\documentclass{aa}

\usepackage{graphics}
\def\mincir{\ \raise -2.truept\hbox{\rlap{\hbox{$\sim$}}\raise5.truept
\hbox{$<$}\ }}
\def\magcir{\ \raise -2.truept\hbox{\rlap{\hbox{$\sim$}}\raise5.truept
\hbox{$>$}\ }}

\begin{document}

\thesaurus{12.03.3} 
\title{The Ly$\alpha$ forest at $1.5 < z < 4$\thanks{Based on public
data released from the UVES commissioning at the VLT/Kueyen
telescope, ESO, Paranal, Chile.}$^{,}$ \thanks{Table A.1, 
Table A.2 and Table A.3
are only available in electronic form at the CDS via anonymous ftp
to cdsarc.u-strasbg.fr (130.79.128.5).}}

\author{Tae-Sun Kim,
\inst{1}
Stefano Cristiani
\inst{2, 3}
\and
Sandro D'Odorico 
\inst{1}}

\offprints{T.-S. Kim}

\institute{European Southern Observatory \\
Karl-Schwarzschild-Strasse 2, D-85748, Garching b.
M\"unchen, Germany\\
e-mail: tkim@eso.org\\
e-mail: sdodoric@eso.org
\and
ST European Coordinating Facility, ESO\\
Karl-Schwarzschild-Strasse 2, D-85748, Garching b.
M\"unchen, Germany\\
e-mail: scristia@eso.org 
\and
Dipartimento di Astronomia dell'Universit\`a di Padova\\
Vicolo dell'Osservatorio 2, I-35122 Padova, Italy
}

\date{Received 22 August, 2000; Accepted}

\maketitle

\begin{abstract}

Using high resolution ($R \sim 45\,000$), high S/N ($\sim$ 20--50)
VLT/UVES data, we have analyzed the Ly$\alpha$ forest 
of 3 QSOs in the neutral hydrogen (\ion{H}{i}) column
density range $N_\ion{H}{i} = 10^{12.5-16} \
\mathrm{cm}^{-2}$ at $1.5 < z < 2.4$. We combined
our results with 
similar high-resolution, high S/N 
data in the literature at $z > 2.4$ to study
the redshift evolution of the Ly$\alpha$ forest at
$1.5 < z < 4$. We have applied two
types of analysis: the traditional Voigt profile fitting and
statistics on the transmitted flux. 
The results from both analyses 
are in good agreement:

\begin{enumerate}

\item The differential column density
distribution function, $f(N_\ion{H}{i})$,
of the Ly$\alpha$ forest shows little evolution 
in the column density range
$N_\ion{H}{i} = 10^{12.5-14} \ \mathrm{cm}^{-2}$,
$f(N_\ion{H}{i}) \propto N_\ion{H}{i}^{-\beta}$,
with $\beta \sim 1.4$--1.5 at $1.5 < z < 4$ and 
with a possible increase of $\beta$ to
$\beta \sim 1.7$ at
$z  < 1.8$.
A flattening of the power law  slope 
at lower column densities at higher $z$ 
can be attributed to more severe line blending. 
A deficiency of lines with
$N_\ion{H}{i} > 10^{14} \ \mathrm{cm}^{-2}$ is more
noticeable at lower $z$ than at higher $z$.
The one-point function and the two-point function of the
flux confirm that strong lines do evolve faster
than weak lines. 

\item The line number density per unit redshift,
$dn/dz$, at
$N_\ion{H}{i} = 10^{13.64-16} \ 
\mathrm{cm}^{-2}$ is well fitted by a single power law, 
$dn/dz \propto (1+z)^{2.19 \pm 0.27}$, 
at $1.5 <z < 4$. In combination with the HST results
from the HST QSO absorption line key project, 
the present data
indicate that a flattening in the number density evolution occurs at
$z \sim 1.2$. The line counts as a function of the
filling factor at the 
transmitted flux $F$ in the range
$0 < F < 0.9$ are constant in the interval $1.5 < z < 4$. 
This suggests
that the Hubble expansion is the main drive governing the
forest evolution at $z > 1.5$ and that the metagalactic UV background
changes more slowly than a QSO-dominated background at $z < 2$.

\item The observed cutoff Doppler parameter
at the fixed column density
$N_\ion{H}{i} = 10^{13.5} \ \mathrm{cm}^{-2}$,
$b_{c, \mathrm{13.5}}$, shows a weak increase with
decreasing $z$, with a possible local $b_{c, \mathrm{13.5}}$
maximum at $z \sim 2.9$.

\item The two-point velocity correlation function and the
step optical depth correlation function
show that the clustering strength increases as $z$ decreases.

\item The evolution of the mean \ion{H}{i} opacity,
$\overline{\tau}_\ion{H}{i}$, is well
approximated by an empirical power law, 
$\overline{\tau}_\ion{H}{i}$ $\propto (1+z)^{3.34 \pm 0.17}$,
at $1.5 < z < 4$.

\item The baryon density, $\Omega_\mathrm{b}$, derived both from 
the mean \ion{H}{i} opacity
and from the one-point function of the flux
is consistent with the hypothesis that most
baryons (over 90\%) 
reside in the forest at $1.5 < z < 4$, with little change in the
contribution to the density, $\Omega$, as a function of $z$.

\end{enumerate}
\keywords{Cosmology: observations -- quasars: Ly$\alpha$ forest --
quasars: individual \object{HE0515--4414}, 
\object{HE2217--2818}, \object{J2233--606},
\object{HS1946+7658}, \object{Q0302--003}, \object{Q0000--263}}
\end{abstract}

%

\section{Introduction}

The Ly$\alpha$ forest imprinted in the spectra of high-$z$ QSOs 
provides a unique and powerful tool to study the distribution/evolution
of baryonic matter and the physical status of the intergalactic 
medium (IGM) over a wide range of $z$ up to $z \sim 6$.
In addition, the Ly$\alpha$ forest can be used to constrain
cosmological parameters, such as the density parameter $\Omega$ and the
baryon density $\Omega_{b}$, providing a test to current
cosmological theories
(Sargent et al. \cite{sar80}; Dav\'e et al. \cite{dav99};
Impey et al. \cite{imp99}; 
Schaye et al. \cite{sch99}; Machacek et al. \cite{mac00}).

\begin{table*}
\caption[]{Observation log}
\label{Tab1}
\begin{tabular}{lcccclc}
\hline
\noalign{\smallskip}
QSO & $B^{a}$ & $z_{\mathrm{em}}$ & Wavelength & Exp. time & 
Observing Date & Comments \\
& (mag) & & (\AA\/) & (sec) & & \\
\noalign{\smallskip}
\hline
\noalign{\smallskip}
\object{HE0515--4414} & 14.9 & 1.719 & 
3050--3860 & 19000 & Dec. 14,
18, 1999 & Reimers et al. (\cite{rei98}) \\
\object{J2233--606} & 17.5 & 2.238 & 3050--3860 & 16200 & Oct. 8-12, 1999 & \\
\object{J2233--606} & & &3770--4980 & 12300 & Oct. 10-16, 1999 & \\
\object{HE2217--2818} & 16.0 & 2.413& 3050--3860 & 16200 & Oct. 5--6, 1999 
& Reimers et al. (\cite{rei96}) \\
\object{HE2217--2818} & & & 3288--4522 & 10800 & Sep. 27--28, 1999 & \\
\noalign{\smallskip}
\hline
\end{tabular}
\begin{list}{}{}
\item[$^{\mathrm{a}}$] Taken from the SIMBAD astronomical database.
\end{list}
\end{table*}

Although detections of \ion{C}{iv} in the
forest clouds suggest that the Ly$\alpha$ forest is closely 
related to galaxies
(Cowie et al. \cite{cow95}; Tytler et al. \cite{tyt95}), 
identification of its optical counterpart at $z < 1$ 
has produced different interpretations: extended haloes of
intervening galaxies (Lanzetta et al. \cite{lan95}; Chen et al.
\cite{che98}) or \ion{H}{i} gas tracing the large-scale distribution
of galaxies and dark matter (Morris et al. \cite{mor93};
Shull et al. \cite{shu96}; Le Brun et al. \cite{le97};
Bowen et al. \cite{bow98}). Despite a lack of positive identification
of optical counterparts of the Ly$\alpha$ forest,
high resolution, high S/N data 
have provided a wealth of information on the cosmic evolution of
the Ly$\alpha$ forest, such as the space density of absorbers,
the distribution of column densities and Doppler widths, 
and the velocity correlation
strengths (Lu et al. \cite{lu96}; 
Cristiani et al. \cite{cri97}; Kim et al. \cite{kim97}).

Up to now, the systematic studies of the Ly$\alpha$ forest at $z < 1.7$
have mostly 
relied on low resolution ($R \sim 1\,300$--$16\,000$) HST 
observations
(Bahcall et al. \cite{bah93}; Weymann et al. \cite{wey98};
Penton et al. \cite{pen00}),
which cannot be properly combined with 
the high-resolution ($R \sim 40\,000$)
ground-based data. 
Here, we present the observations of the Ly$\alpha$
forest at $1.5 < z < 2.4$ using the high resolution ($R \sim 45\,000$),
high S/N ($\sim$ 20--50) VLT/UVES commissioning data on three QSOs. These
observations
take advantage of the high UV sensitivity of UVES
(D'Odorico et al. \cite{dor00}). Combining these data with similar 
Keck\thinspace I/HIRES results at $z > 2.5$ from the literature, 
we address the $z$-evolution of the Ly$\alpha$ forest at $1.5 < z < 4$ 
as well as the physical properties of the Ly$\alpha$ forest having 
$N_\ion{H}{i} = 10^{12.5-16} \, \mathrm{cm}^{-2}$.
In addition, when appropriate, we also compare our results
with HST observations at $z < 1.7$.
In Sect. 2, we describe the UVES observations and data
reduction. In Sect. 3, we describe the conventional 
Voigt profile fitting
technique and its application to the UVES spectra of the Ly$\alpha$
forest in this study. In Sect. 4, we analyze the 
line sample obtained from 
the Voigt profile fitting. 
In Sect. 5, we show an analysis based on the transmitted flux or its
optical depth, 
which supplements the Voigt profile fitting analysis
in Sect. 4. This parallel analysis has
the advantage of including absorptions with low optical depths
which are usually excluded from the Voigt profile
fitting analysis. It also gives a more robust comparison with 
numerical simulations and other observations at similar S/N and resolution.
We discuss our overall results in Sect. 6 and 
the conclusions are summarized in Sect. 7. 
In this study, all the quoted uncertainties are $1\sigma$ errors.

\section{Observations and data reduction}

The data presented here
were obtained during the Commissioning I and II of UVES as
a test of the instrument capability in the UV region and have been
released by ESO for public use. 
The properties
of the spectrograph and of its detectors are described in Dekker et al.
(\cite{dek00}).
Among the QSOs observed with UVES,
we selected three QSOs for the Ly$\alpha$ forest study at $z \sim 2$:
\object{HE0515--4414}, \object{J2233--606} and 
\object{HE2217--2818}. 

Complete wavelength coverage
from the UV atmospheric cutoff $\sim 3050$ \AA\/ 
to $\sim$ 5000 \AA\/ was obtained for the three
QSOs with two setups which use dichroic beam splitters to feed the blue
and red arm of the spectrograph in parallel. In this paper we discuss only
the Ly$\alpha$ forest observations
which were recorded in the blue arm of the spectrograph.
The pixel size in the direction of the dispersion corresponds to 
0.25 arcsec in the
blue arm and the slit width was used 
typically 0.8--0.9 arcsec (the narrower slit being used for the 
brighter target \object{HE0515-4414}).
The resolving power, as measured from 
several isolated Th-Ar lines distributed over the spectrum and
extracted in the same way as the object spectra,
is $\sim 45\,000$ in the regions of interest.
Table~\ref{Tab1} lists the observation log
and the magnitude of the observed QSOs. The exposure times
are the sum of individual integrations ranging from $2\,700$ 
to $5\,000$ seconds.

The UVES data were reduced with the ESO-maintained MIDAS ECHELLE/UVES 
package. The individual frames were bias-subtracted 
and flat-fielded. The cosmic rays were flagged using a median
filter. The sky-subtracted spectra were then optimally extracted,
wavelength-calibrated, and merged.
The wavelength calibration was 
checked with the sky lines such as [\ion{O}{i}] 5577.338 \AA\/,
\ion{Na}{i}
5989.953 \AA\/, and OH bands (Osterbrock et al. \cite{ost96}). The
typical uncertainty in wavelength is $\sim 0.01$ \AA\/. 
The wavelengths in the final spectra
are vacuum heliocentric. 
The individually reduced spectra were
combined with weighting corresponding to their S/N and resampled with
a 0.05 \AA\/ bin.
The S/N varies across the
spectrum, increasing towards longer wavelengths for a given
instrumental configuration.
The typical S/N per pixel is $\sim$ 20--50 for \object{HE0515--4414}
at 3090--3260 \AA\/,
$\sim$ 25--40 for \object{J2233--606} at 3400--3850 \AA\/ and
$\sim$ 45--50 for \object{HE2217--2818} at 3550--4050 \AA\/.

The combined
spectra were then normalized locally using a 5th order polynomial
fit.
There is no optimal method to determine the real underlying continuum of
high-$z$ QSOs at wavelengths blueward of the Ly$\alpha$ emission due
to high numbers of Ly$\alpha$ absorptions.
The normalization of the spectra introduces the
largest uncertainty in the study of weak forest lines. However,
considering the high resolution of our data and the 
relatively low number density
of the forest at $z \sim 2$, the continuum uncertainty should be
considerably less than 10\%.

\section{The Voigt profile fitting}

Conventionally, the Ly$\alpha$ forest has been thought of as
originating in discrete clouds
and has thus been analyzed as a collection of
{\it individual lines} whose characteristics can be
obtained by fitting the Voigt profiles.
From the line fitting, three parameters are derived:
the redshift of an absorption line, $z$, its Doppler parameter,
$b$ (if the line is broadened thermally, 
the $b$ parameter gives the thermal temperature of a gas,
$b \equiv \sqrt{2kT/m_{p}}$, where $T$ is the gas temperature,
$m_{p}$ is the proton mass, and $k$ is the Boltzmann
constant), 
and its \ion{H}{i} column density, $N_\ion{H}{i}$.

We used Carswell's VPFIT program (Carswell et al.:
http://www.ast.cam.ac.uk/$\sim$rfc/vpfit.html) to fit the absorption
lines.
For a selected wavelength region, VPFIT adjusts the
initial guess solution to minimize the $\chi^{2}$ between the
data and the fit. 
We have chosen the reduced $\chi^{2}$ threshold
for an acceptable fit to be $1.3$ and
we add more components if $\chi^{2} \ge 1.3$. Even though
the adopted threshold is somewhat arbitrary,
a difference between the $\chi^{2} \le 1.3$ threshold and 
the $\chi^{2} \le 1.1$ threshold is negligible when
line blending is not severe, like at $z \sim 2$.
Note that there is no unique solution for 
the Voigt profile fitting (cf. Kirkman \& Tytler \cite{kir97}).
In particular, for high S/N data, absorption profiles show various
degrees of departure from the Voigt profile
(cf. Rauch \cite{rau96}; Outram et al. \cite{out99b}).
This departure can be
fitted by adding one or two physically improbable 
narrow, weak lines, which results in
overfitting of line profiles (see Sect. 5 for further discussion).

In this study, metal lines were excluded as follows:
When isolated metal lines were identified, these portions of the
spectrum were substituted by a mean normalized flux of 1
with 
noise similar to nearby spectral regions. When metal lines were
embedded in a complex of \ion{H}{i} lines, the complex was fitted
with Voigt profiles and the 
contribution from the metal lines was subtracted from the
profile of the complex. 
Although metal lines were searched for
thoroughly, it is possible that some unidentified metal lines are
present in the \ion{H}{i} line lists from VPFIT. In most cases,
absorption lines with $b < 15 \ \mathrm{km s}^{-1}$
can be attributed to metal lines (Rauch et al. \cite{rau97}).
In our line lists, these narrow lines are less than 5\% of the 
absorption lines not identified as metal lines.
Therefore, including these 
narrow lines does not change our conclusions significantly.

We only consider the regions of a spectrum
between the QSO's Ly$\alpha$ and Ly$\beta$ emission lines to
avoid confusion with the Ly$\beta$ forest. In addition,
we exclude the regions close to the
QSO's emission redshift to avoid the proximity effect.
For \object{HE0515--4414}, we exclude a region of $4\,500$ km s$^{-1}$,
while for \object{J2233--606} and \object{HE2217--2818}
we exclude a region of $\sim 7\,000$ km s$^{-1}$.

\begin{figure*}
\resizebox{\hsize}{!}{\includegraphics{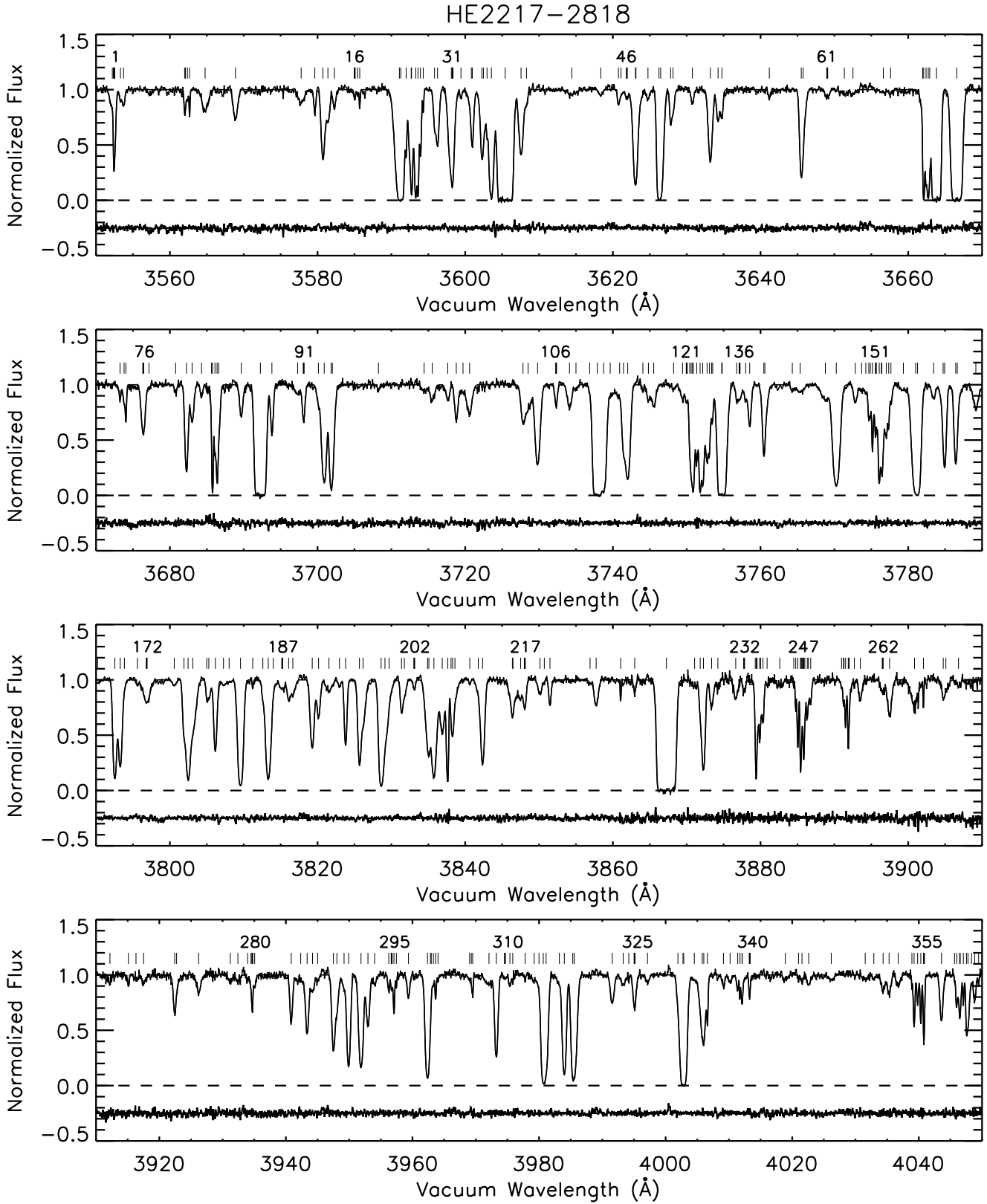}}
\caption{The spectrum of \object{HE2217--2818} superposed with the Voigt
profile fit.
The residuals (the differences between the observed
and the fitted flux) shown in the bottom part of each panel are shifted
by $-0.25$.}
\label{fig_he22_norm1}
\end{figure*}

\begin{figure*}
\resizebox{\hsize}{!}{\includegraphics{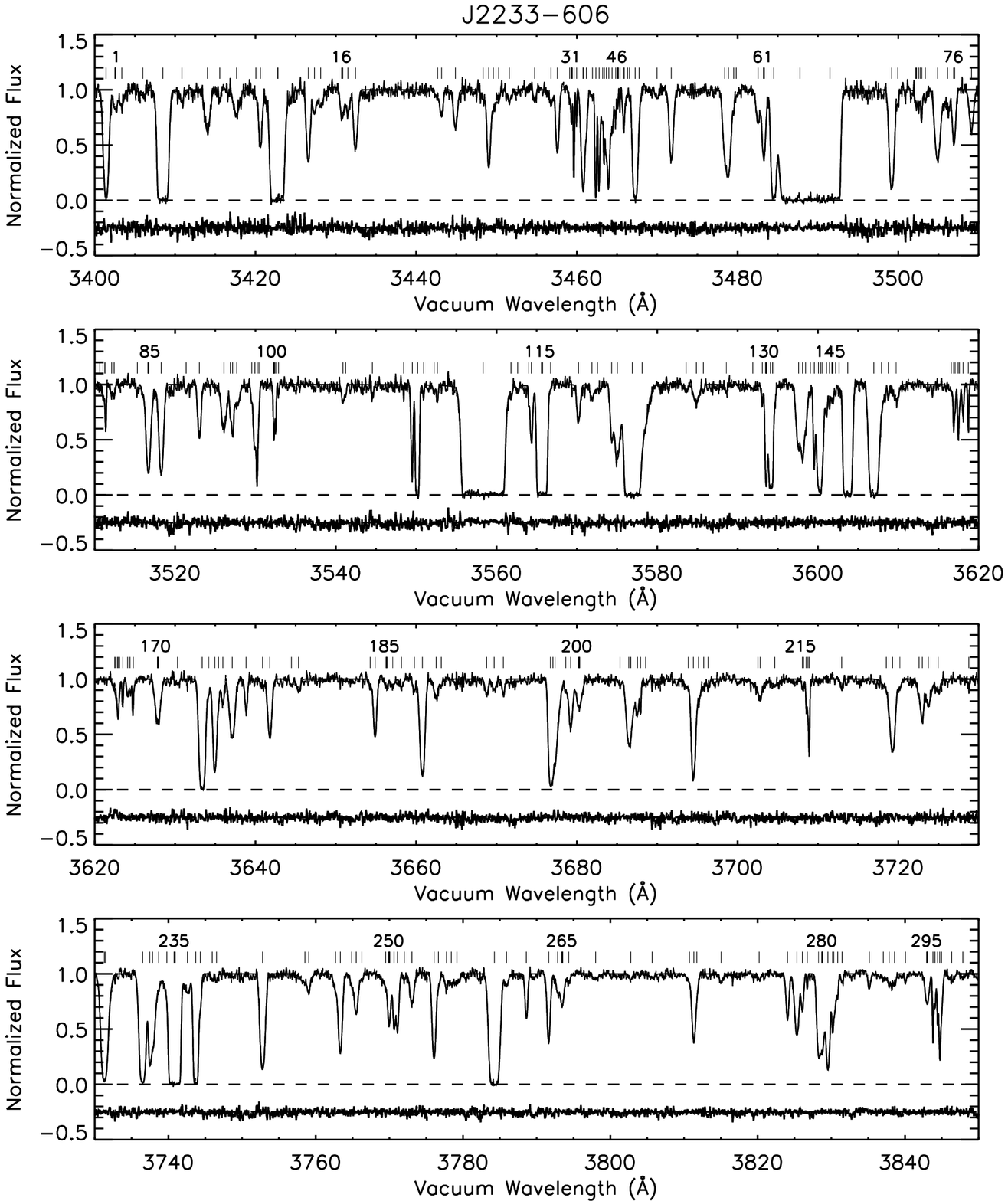}}
\caption{The spectrum of \object{J2233--606} superposed with the Voigt
profile fit.
The residuals (the differences between the observed
and the fitted flux) shown in the bottom part of each panel are shifted
by $-0.25$.}
\label{fig_j22_norm1}
\end{figure*}

\begin{figure*}[t]
\resizebox{\hsize}{!}{\includegraphics{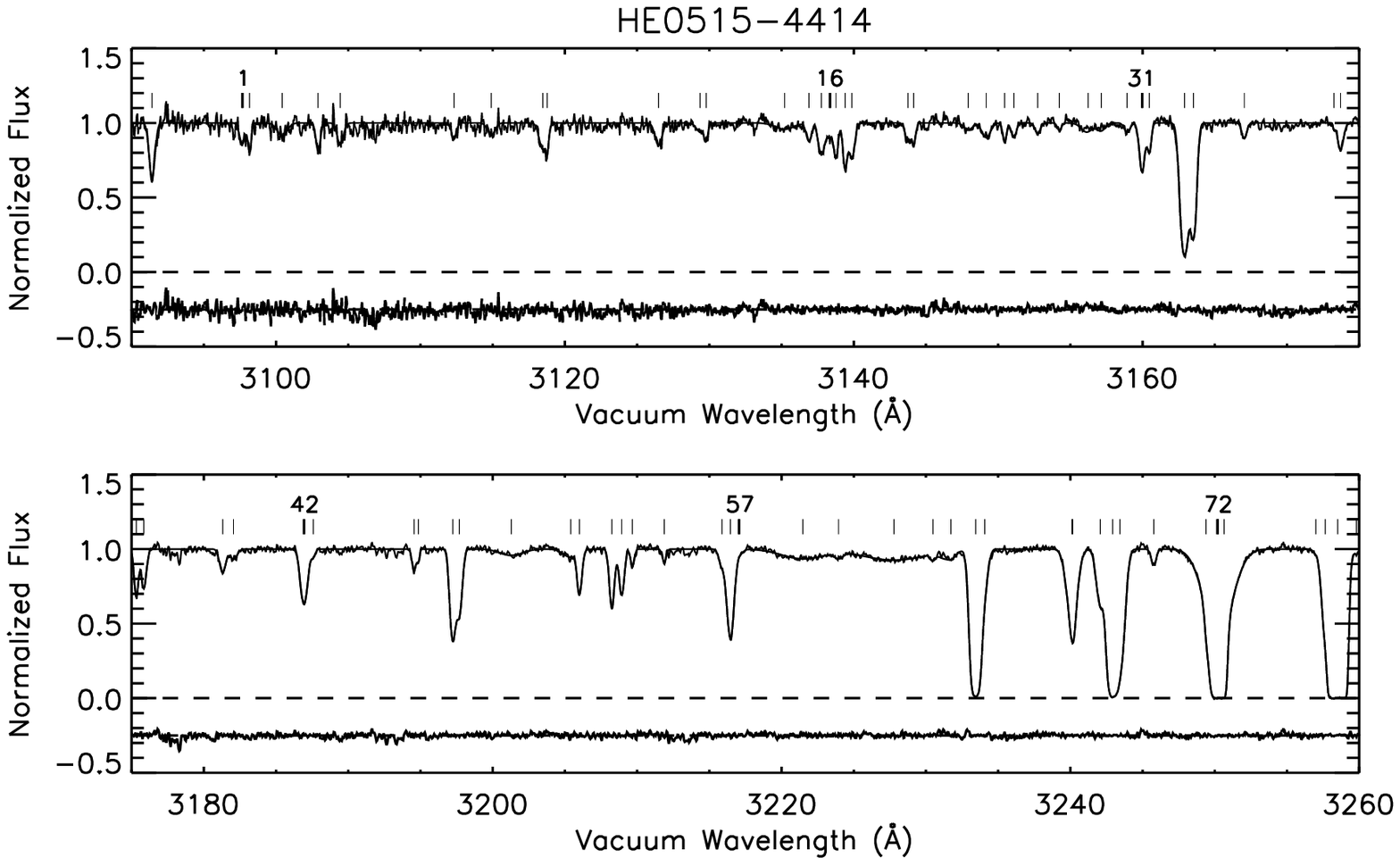}}
\caption{The spectrum of \object{HE0515--4414} superposed with the Voigt
profile fit.
The residuals (the differences between the observed
and the fitted flux) shown in the bottom part of each panel are shifted
by $-0.25$.}
\label{fig_he05_norm1}
\end{figure*}

Figs.~\ref{fig_he22_norm1}, \ref{fig_j22_norm1} and \ref{fig_he05_norm1} 
show the
spectra of \object{HE2217--2818}, \object{J2233--606} and
\object{HE0515--4414}, respectively,
superposed with their Voigt profile fit
(the fitted line lists [Tables A.1--A.3] from VPFIT with
their errors are only
electronically published). The tick marks indicate the center of
the lines fitted with VPFIT and the numbers above the bold tick marks
indicate the number of the fitted line in the line lists, which starts from 0.
From the residuals between the observed and the fitted spectra,
the variation of S/N across the spectra is easily recognizable.
Due to the limited S/N in the data,
lines with $N_\ion{H}{i} < 10^{12.5} \ \mathrm{cm}^{-2}$
become confused with noise. Therefore, we restricted
our analysis to $N_\ion{H}{i} = 10^{12.5-16.0} \ \mathrm{cm}^{-2}$.
We included all the forest lines regardless of the existence of
associated metals because results from the HST observations 
are in general 
based on the equivalent width of the forest lines, not on
the existence of metals (Weymann et al. \cite{wey98};
Savaglio et al. \cite{sav99}; Penton et al. \cite{pen00}). 
In addition, the detection of metal lines in 
the forest at $N_\ion{H}{i} \ge
10^{14} \ \mathrm{cm}^{-2}$ and at $z \sim 3$ make it 
unclear whether there
is a spread in the metallicity in the Ly$\alpha$ forest or if there exists
two different populations, such as a metal-free
forest and a metal-contaminated Ly$\alpha$ forest (Songaila
\cite{son98}; Ellison et al. \cite{ell99}).

\section{The Voigt profile analysis of the Ly$\alpha$ forest}

In addition to the three QSOs observed with UVES,
we have used the published line lists
of three QSOs at higher redshift observed
with similar resolution and S/N.
Table~\ref{Tab2} lists all the analyzed QSOs
with their properties and the relevant references.
We have avoided a region close to 
a damped Ly$\alpha$ system in the spectrum of \object{Q0000--263}
and a lower S/N region in the spectrum 
of \object{HS1946+7658}. The spectrum of \object{Q0302--003} 
does not include the region of the known void at $z \sim 3.17$
(cf. Dobrzycki \& Bechtold \cite{dob91}).
The fitted line parameters with the associated errors 
of \object{HS1946+7658} and \object{Q0000--263} 
were generated by VPFIT with a $\chi^2$ threshold of
$\chi^2 \le 1.2$ and $\chi^2 \le 1.1$, respectively.
The line list of \object{Q0302--003}
was generated by an automatized version of 
the Voigt profile fitting program by Hu et al. (\cite{hu95})
with $\chi^2 \sim 1$ and the errors associated with the fitted parameters
are not published.

The results of profile fitting are known to be sensitive to the data 
quality as well as to the characteristics of the fitting program.
As a consequence, comparing line lists obtained with different criteria
is not usually straightforward.
Due to the use of a different fitting program,
the line list of \object{Q0302--003} at $z \sim 2.9$ should be
treated with caution
when combined with other line lists. A systematic difference in
$b$ and $N_\ion{H}{i}$ from VPFIT can introduce a slightly different
behavior of the Ly$\alpha$ forest at $z \sim 2.9$. While the
difference would not change the study of the line number
density or the correlation function significantly, it can affect
the determination of a lower cutoff $b$ envelope in 
the $N_\ion{H}{i}$-$b$ diagrams.
Furthermore, the six QSOs in Table~\ref{Tab2}
cover the Ly$\alpha$ forest at $1.54 < z_\mathrm{Ly\alpha}
< 4.0$ with a fairly regular spacing.
There is very little overlap between the Ly$\alpha$ 
forests of the different QSOs and the effects
of cosmic variance in the individual lines of sight might be important.

\begin{table}
\caption[]{Analyzed QSOs}
\label{Tab2}
\begin{tabular}{lcccc}
\hline
\noalign{\smallskip}
QSO & $z_\mathrm{em}$ & $\lambda\lambda$ & $z_\mathrm{Ly\alpha}$ &
$dX^{\mathrm{a}}$ \\
\noalign{\smallskip}
\hline
\noalign{\smallskip}
\object{HE0515--4414} & 1.719 & 3090--3260 & 1.54--1.68 & 0.365 \\
\object{J2233--606}$^{\mathrm{b}}$ & 2.238 & 3400--3850 & 1.80--2.17 & 1.104 \\
\object{HE2217--2818} & 2.413 & 3550--4050 & 1.92--2.33 & 1.286 \\
\object{HS1946+7658}$^{\mathrm{c}}$
& 3.051 & 4252--4635 & 2.50--2.81 & 1.157 \\
\object{Q0302--003}$^{\mathrm{d}}$
& 3.290 & 4410--5000 & 2.63--3.11 & 1.878 \\
\object{Q0000--263}$^{\mathrm{e}}$
& 4.127 & 5450--6100 & 3.48--4.02 & 2.540 \\
\noalign{\smallskip}
\hline
\end{tabular}
\begin{list}{}{}
\item[$^{\mathrm{a}}$] For $q_\mathrm{0}=0$.
\item[$^{\mathrm{b}}$] See also Cristiani \& D'Odorico (\cite{cri00})
\item[$^{\mathrm{c}}$] Kirkman \& Tytler (\cite{kir97})
\item[$^{\mathrm{d}}$] Hu et al. (\cite{hu95})
\item[$^{\mathrm{e}}$] Lu et al. (\cite{lu96})
\end{list}
\end{table}

\subsection{The differential density distribution function}
The differential density distribution function,
$f(N_\ion{H}{i})$,
is defined as the number of absorption lines per unit absorption
distance path
and per unit column density as a function of $N_\ion{H}{i}$
(equivalent to the luminosity function of galaxies). 
The absorption distance path $X(z)$ is defined as $X(z)
\equiv {1\over 2}
[(1+z)^{2} -1]$ for $q_\mathrm{o} = 0$ or as $X(z) \equiv {2\over 3}
[(1+z)^{3/2} -1]$ for $q_\mathrm{o} = 0.5$.
We used $q_\mathrm{o} = 0$ for $dX$ to compare our
$f(N_\ion{H}{i})$ with
the published $f(N_\ion{H}{i})$ from the literature
(Table~\ref{Tab2}
lists the values of $dX$ for $q_\mathrm{o} = 0$).
Empirically, $f(N_\ion{H}{i})$ is fitted by a power law:
$f(N_\ion{H}{i}) = A \, N_\ion{H}{i}^{-\beta}$.

\begin{figure}
\resizebox{\hsize}{!}{\includegraphics{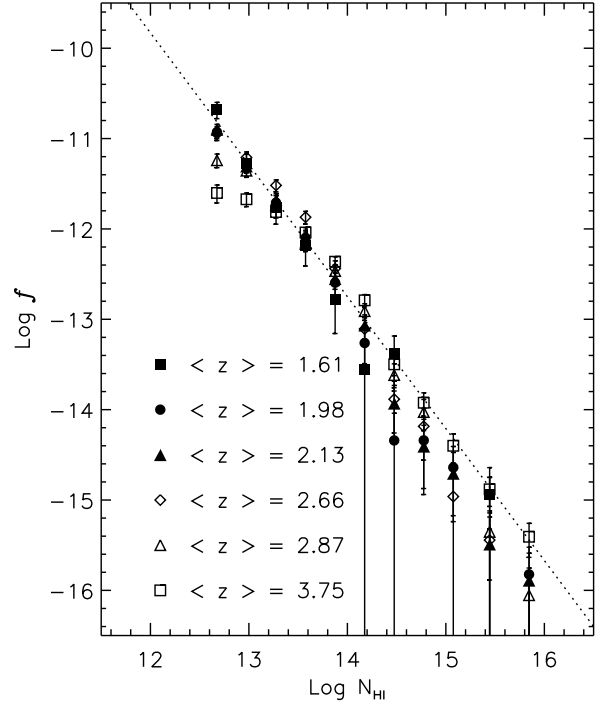}}
\caption{The differential density distribution function as a
function of $\log N_\ion{H}{i}$ without the incompleteness correction due to
line blending and limited S/N. The vertical error bars represent
the $1\sigma$ Poisson errors.}
\label{fig_ddf}
\end{figure}

Fig.~\ref{fig_ddf} shows the {\it observed} $\log f(N_\ion{H}{i})$
as a function of $\log N_\ion{H}{i}$ for different redshifts. 
The dotted line represents
the incompleteness-corrected $f(N_\ion{H}{i})$ at
$z \sim 2.8$ from Hu et al. (\cite{hu95}),
i.e. $f(N_\ion{H}{i}) =4.9 \times 10^{7} \, N_\ion{H}{i}^{-1.46}$.
Note that the {\it apparent} flattening of the slope towards 
lower column densities in the observed 
$\log N_\ion{H}{i}$-$\log f(N_\ion{H}{i})$
diagram is caused by line blending and limited S/N, i.e. incompleteness, which
becomes more severe at higher $z$.
For incompleteness-corrected $f(N_\ion{H}{i})$
at $z > 2.5$ (Hu et al. \cite{hu95}; Lu et al. \cite{lu96};
Kim et al. \cite{kim97}), this apparent flattening disappears.
The incompleteness-corrected $f(N_\ion{H}{i})$ at $z \sim 2.8$ 
over $N_\ion{H}{i}= 10^{12.5-16} \ \mathrm{cm}^{-2}$
is similar to the incompleteness-corrected $f(N_\ion{H}{i})$ 
at $z \sim 3.7$ (Lu et al. \cite{lu96}) and
at $z \sim 3.2$ (Kim et al. \cite{kim97}) over the same column density
range.
The amount of 
incompleteness extrapolated from at $z > 2.5$ (Hu et al. \cite{hu95}; 
Lu et al. \cite{lu96};
Kim et al. \cite{kim97}; Kirkman \& Tytler \cite{kir97})
becomes negligible at $z < 2.4$ and we assume the {\it observed}
$f(N_\ion{H}{i})$ as representative of the {\it actual}
$f(N_\ion{H}{i})$ at $z < 2.4$. 

In the column density range 
$N_\ion{H}{i} = 10^{12.5-14} \ \mathrm{cm}^{-2}$,
the observed $f(N_\ion{H}{i})$ at $z < 2.4$ is in
good agreement for the different QSOs
and also agrees
with the incompleteness-corrected $f(N_\ion{H}{i})$ at 
$2.6< z < 4.0$.
This suggests that there is very little
evolution in $f(N_\ion{H}{i})$ in the interval $1.5 <z < 4$
for forest lines with 
$N_\ion{H}{i} = 10^{12.5-14} \ \mathrm{cm}^{-2}$.
At $N_\ion{H}{i} = 10^{14-16} \ \mathrm{cm}^{-2}$, $f(N_\ion{H}{i})$
shows differences at different $z$. Kim et al. (\cite{kim97}) noted
that at lower $z$, $f(N_\ion{H}{i})$ starts to deviate from a single power
law for $N_\ion{H}{i} \magcir 10^{14} \
\mathrm{cm}^{-2}$ and that the column density at which the deviation 
from a single power-law starts 
decreases as $z$ decreases. The deviation from the single power-law
in $f(N_\ion{H}{i})$ is evident in Fig.~\ref{fig_ddf}. 
While the forest at $z \sim 3.7$ is still well approximated by a single
power-law over $N_\ion{H}{i} = 10^{14-16} \ \mathrm{cm}^{-2}$, 
the forest at $z<2.4$ starts to deviate from the power law
at $N_\ion{H}{i} \ge 10^{14.1} \ \mathrm{cm}^{-2}$ with a 
decreasing number of lines at 
$N_\ion{H}{i} = 10^{14-16} \ \mathrm{cm}^{-2}$.

Table~\ref{Tab3} lists the parameters of a 
maximum-likelihood power-law fit to various column density ranges.
These column density ranges are selected for comparison with
the previous observational results of Kim et al. (\cite{kim97})
and Penton et al. (\cite{pen00}) and with simulations of
Zhang et al. (\cite{zha98}) and Machacek et al. (\cite{mac00}).
At $z \sim 2.1$, the slope $\beta$ is
approximately 1.4 in the interval $N_\ion{H}{i} = 10^{12.5-14} \
\mathrm{cm}^{-2}$ and 1.68 in the interval $N_\ion{H}{i} = 10^{14-16} \
\mathrm{cm}^{-2}$, i.e. the slope is steeper for
higher column density clouds. At $z \sim 1.61$,
the slopes $\beta \sim 1.70$--1.72 are steeper
for both column density ranges.
This indicates that the slope of
$f(N_\ion{H}{i})$ increases from $z \sim 2.1$ to $z \sim 1.6$. 
Assuming a curve of growth with $b = 25 \pm 5$ km s$^{-1}$, 
Penton et al. (\cite{pen00}) found that the slope of $f(N_\ion{H}{i})$ 
at $z \sim 0.036$ over $N_\ion{H}{i} = 10^{12.5-14} \ \mathrm{cm}^{-2}$
and over $N_\ion{H}{i} = 10^{14-16} \ \mathrm{cm}^{-2}$ is 
$\beta = 1.72 \pm 0.06$
and $\beta = 1.43 \pm 0.35$, respectively.
The slopes over $N_\ion{H}{i} = 10^{12.5-14} \ \mathrm{cm}^{-2}$
are steeper at $z \sim 0.036$ and at $z \sim 1.6$ than
at $z > 1.8$, and suggest that the incompleteness correction at
$z > 1.8$ might be underestimated or that the slope becomes
intrinsically steeper at $z < 1.8$. 
The slopes over $N_\ion{H}{i} = 10^{12.8-14} \ \mathrm{cm}^{-2}$
and $N_\ion{H}{i} = 10^{14-16} \ \mathrm{cm}^{-2}$ are
in agreement with the ones found by Kim et al. (\cite{kim97})
at $z \sim 2.3$. However,
our measurement of 
$\beta = 1.48 \pm 0.15$ at $z \sim 2.1$ (only from 
\object{HE2217--2818} and \object{J2233--606}) 
over $N_\ion{H}{i} = 10^{13.1-14.5} \ \mathrm{cm}^{-2}$ 
is lower than the previous determination of 
$\beta = 1.79 \pm 0.10$ over the 
same column density range 
at $z \sim 1.85$ by Kulkarni et al. (\cite{kul96}). 

While these observed $\beta$ values at $1.5 < z < 2.4$ can be
obtained with 
semi-analytic models by Hui et al. (\cite{hui97b}), 
they are lower than the values predicted
from numerical simulations 
(Zhang et al. \cite{zha98}; Machacek et al. \cite{mac00}),
by more than  
$2\sigma$.
The slope depends on the
amplitude of the power spectrum and models with less power
produce steeper slopes. Thus, the steeper slopes from the simulations
by Zhang et al. (\cite{zha98}) and Machacek et al. (\cite{mac00})
suggest that their index for the power spectrum, $n_{p}=1$,
might be smaller than the {\it actual} index of the power spectrum.

\begin{table*}
\caption[]{The power-law fit of the distribution functions,
$f(N_\ion{H}{i}) = A\,N_\ion{H}{i}^{-\beta}$} 
\label{Tab3}
\begin{tabular}{cccccccccccc}
\hline
\noalign{\smallskip}
& \multicolumn{2}{c}{$N_\ion{H}{i} = 10^{12.5-14.0} \ \mathrm{cm}^{-2}$}&
&
\multicolumn{2}{c}{$N_\ion{H}{i} = 10^{14.0-16.0} \ \mathrm{cm}^{-2}$} &
&
\multicolumn{2}{c}{$N_\ion{H}{i} = 10^{12.8-14.3} \ \mathrm{cm}^{-2}$} &
&
\multicolumn{2}{c}{$N_\ion{H}{i} = 10^{12.5-16.0} \ \mathrm{cm}^{-2}$} \\
\\[-2ex]
\cline{2-3} \cline{5-6} \cline{8-9} \cline{11-12}\\[-2ex]
$z$ & $\log A$ & $\beta$ & &
$\log A$ & $\beta$ & &
$\log A$ & $\beta$ & &
$\log A$ & $\beta$ \\
\noalign{\smallskip}
\hline
\noalign{\smallskip}
1.61 & $10.65 \pm 0.86$ & $1.72 \pm 0.16$ & 
& $10.38 \pm 0.35$ 
 & $1.70 \pm 0.42$ & &
 $10.85 \pm 0.74$ & $1.74 \pm 0.21$ & 
 & $10.47 \pm 0.88$ & $1.71 \pm 0.10$ \\
1.98 & $6.19 \pm 1.07$ & $1.35 \pm 0.09$ & 
 & $6.49 \pm 0.59$ & $1.40 \pm 0.21$ & &
 $8.02 \pm 1.00$ & $1.49 \pm 0.11$ & 
 & $8.66 \pm 1.09$ & $1.54 \pm 0.05$ \\
2.13 & $6.66 \pm 1.11$ & $1.38 \pm 0.08$ & & 
  $14.48 \pm 0.65$ & $1.94 \pm 0.24$ & &
  $7.80 \pm 1.05$ & $1.46 \pm 0.10$ & & 
  $9.05 \pm 1.14$ & $1.56 \pm 0.05$ \\
$1.94^{\mathrm{a}}$ &  $6.97 \pm 1.28$ & $1.41 \pm 0.06$ & & 
  $10.43 \pm 0.80$ & $1.68 \pm 0.15$ & &
  $8.16 \pm 1.21$ & $1.50 \pm 0.07$ & 
 &  $9.02 \pm 1.30$ & $1.57 \pm 0.03$ \\
\noalign{\smallskip}
\hline
\end{tabular}
\begin{list}{}{}
\item[$^{\mathrm{a}}$] $1.54 \le z \le 2.33$.
\end{list}
\end{table*}

\subsection{The evolution of the line number density}

The line number density per unit redshift is defined as $dn/dz =
(dn/dz)_\mathrm{0} (1+z)^{\gamma}$, where 
$(dn/dz)_\mathrm{0}$ is the
local comoving number density of the forest. For a non-evolving
population in the standard Friedmann universe with the cosmological constant
$\Lambda = 0$\footnote{Recent measurements from high-z supernovae 
favor the non-zero cosmological constant (Perlmutter et al. 
\cite{per99}). When we use the mass density $\Omega_{\mathrm{m}} \sim
0.3$ and the cosmological constant energy density $\Omega_{\Lambda}
\sim 0.7$ for the flat universe as the results from the supernova
study favor (Perlmutter et
al. \cite{per99}), the line number density 
for the non-evolving
forest can still be approximated by a single power-law $dn/dz \propto
(1+z)^{\sim 0.71}$ at $0 < z < 4.5$
with a slight steepening at $z < 1$: At $z > 1$, $dn/dz \propto
(1+z)^{\sim 0.59}$, while $z < 1$, $dn/dz \propto (1+z)^{\sim 1.15}$.},
$\gamma = 1$ and 0.5 for $q_\mathrm{0}=0$ and 0.5,
respectively. In practice, the measured $\gamma$ is dependent on the
chosen column density interval, the redshift and the spectral
resolution. Therefore, comparisons between individual studies
are complicated (Kim et al. \cite{kim97}).

\begin{figure}
\resizebox{\hsize}{!}{\includegraphics{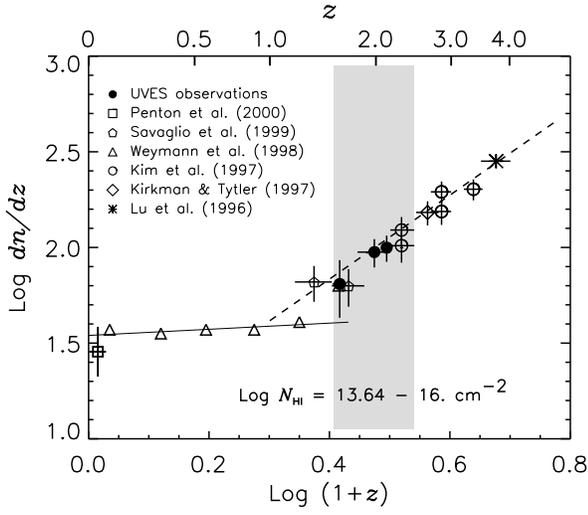}}
\caption{The number density evolution of the Ly$\alpha$ forest.
The column density range
$N_\ion{H}{i} = 10^{13.64 - 16} \ {\rm cm}^{-2}$ has been chosen
to allow a comparison with the HST results 
from Weymann et al. (\cite{wey98}),
which are shown as open triangles.
Filled symbols are estimated from \object{HE0515--4414}
at $<\!z\!> \ = 1.61$, from \object{J2233--606} at
$<\!z\!> \ = 1.98$ and from \object{HE2217--2818} at
$<\!z\!> \ = 2.13$, respectively. 
The star, open circles, and the diamond are taken
from the HIRES data at similar resolutions by Lu et al.
(\cite{lu96}),
Kim et al. (\cite{kim97}), and Kirkman \& Tytler (\cite{kir97}), 
respectively.
The square and pentagons are taken from Penton et al.
(\cite{pen00}) and Savaglio et al. (\cite{sav99}), respectively,
over $N_\ion{H}{i} \ge 10^{14} \ {\rm cm}^{-2}$. 
Horizontal solid lines represent the $z$ interval over which
the number density was estimated. Vertical solid lines represent
the $1\sigma$ Poisson errors.
The shaded area is the $z$ range where UVES is extremely
sensitive.
The long-dashed line is the maximum likelihood fit to the UVES and
the HIRES data at $z > 1.5$.
The UVES observations indicate that the number density evolution 
of the Ly$\alpha$ forest at $z > 2.4$ continues at least
down to $z \sim 1.5$ and that a slope change occurs at $z \sim 1.2$.}
\label{fig_dndz1}
\end{figure}

Fig.~\ref{fig_dndz1} shows the number density evolution of the
Ly$\alpha$ forest in the interval 
$N_\ion{H}{i} = 10^{13.64 - 16} \ {\rm cm}^{-2}$.
This range has been chosen to allow a
comparison with the HST results from the HST QSO absorption 
line Key Project at $z < 1.5$ from Weymann et al.
(\cite{wey98}), for which a threshold in
the equivalent width of 0.24 \AA\/ was adopted.
We assumed the conversion between the equivalent width and 
the column density to be $N_\ion{H}{i} = 1.33 \times 10^{20} 
W/\lambda_\mathrm{0}^{2}f$, where $W$ is the equivalent width in
angstrom, $\lambda_\mathrm{0}$ is the wavelength of Ly$\alpha$ in
angstrom, $f$ is the oscillator strength of Ly$\alpha$
(Cowie \& Songaila \cite{cow86}). 
The value of the square (Penton et al. \cite{pen00}) 
was estimated under the assumption of $b=25$ km s$^{-1}$ from
the equivalent widths (corresponding to the column density range
$N_\ion{H}{i} \ge 10^{14} \ \mathrm{cm}^{-2}$) and is lower than
the extrapolated $dn/dz$ at $z \sim 0.04$ from the Weymann et al.
results, but within the error bar.  
Pentagons (Savaglio et al. \cite{sav99} from the line fitting 
analysis) also correspond to the column density range
$N_\ion{H}{i} \ge 10^{14} \ \mathrm{cm}^{-2}$.
Note that $N_\ion{H}{i}$ from $W$ depends on an assumed $b$ parameter,
resolution and S/N.
Also note that 
including lines with $N_\ion{H}{i} \ge 10^{16} \mathrm{cm}^{-2}$
from line fitting analyses 
introduces a further uncertainty on the line counting since different
programs deblend completely saturated lines differently, 
resulting in different
numbers of lines for the same saturated lines.

The long-dashed line is the maximum-likelihood fit 
to the UVES and the HIRES data at $z > 1.5$:
$dn/dz = (9.06 \pm 0.40) \,(1+z)^{2.19 \pm 0.27}$.
This $\gamma$ is lower than previously reported
$\gamma \sim 2.75 \pm 0.30$ (Lu et al. \cite{lu91}; Kim et al.
\cite{kim97}). 
This slope is steeper than the expected values for the non-evolving
forest for a universe with
$\Omega_\Lambda = 0.7$, $\Omega_\mathrm{m} = 0.3$ and
$\Omega = 1$.
These results suggest that the Ly$\alpha$ forest at 
$N_\ion{H}{i} = 10^{13.64-16} \mathrm{cm}^{-2}$ evolves and that
its evolution 
slows down as $z$ decreases. 
Interestingly, the HST data point at $<\!z\!> \ =1.6$ (the open
triangle at the boundary of the shaded area), which has been measured
in the line-of-sight to the QSO \object{UM 18} and was
suggested to be an outlier by Weymann et al. (\cite{wey98}), is
now in good agreement with the extrapolated fit from higher $z$.

Despite the different line counting methods between the HST observations
(based on the equivalent width) 
and the high-resolution observations
(based on the profile fitting), a change of the slope in the Ly$\alpha$
number density does seem to be real. The
UVES observations suggest that the slow-down in the evolution does occur 
at $z \sim 1.2$, rather than at $z \sim 1.7$ as previously suggested
(Impey et al. \cite{imp96}; Weymann et al. \cite{wey98}),
although the different methods of line counting at higher and lower
$z$ make it a little uncertain.
At least, down to $z \sim 1.5$, the number density of the forest evolves as 
at higher $z$, which suggests that any major drive governing the
forest evolution at $z > 2$ continues to dominate the forest evolution
down to $z \sim 1.5$. Since the Hubble expansion is the main 
drive
governing the forest evolution at $z > 2$ (Miralda-Escud\'e et al.
\cite{mir96}), the continuously decreasing number density of
the forest down to $z \sim 1.5$ implies that the Hubble expansion
continues to dominate the forest evolution down to $z \sim 1.5$.

\begin{figure}
\resizebox{\hsize}{!}{\includegraphics{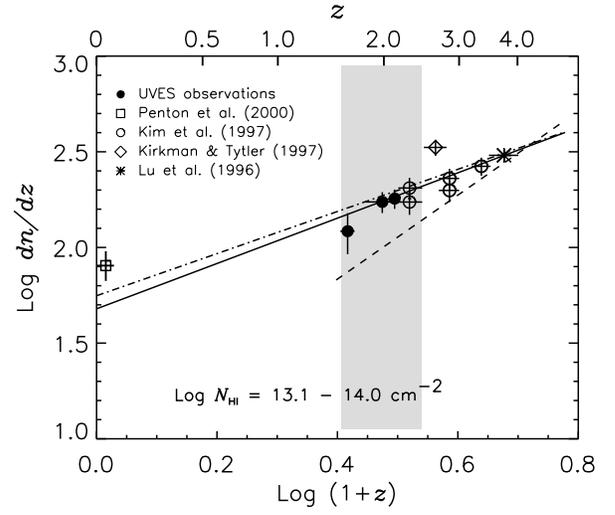}}
\caption{The number density evolution of the Ly$\alpha$ forest at
the $N_\ion{H}{i}$ range $N_\ion{H}{i} = 10^{13.1-14.0} \
\mathrm{cm}^{-2}$. Symbols are the same as in Fig.~\ref{fig_dndz1}.
The dot-dashed line (solid line) 
is the maximum likelihood fit to the lower column
density forest including the $<\!z\!> \, = 2.66$ forest (excluding
the $<\!z\!> \, = 2.66$ forest), 
while the dashed line represents the maximum
likelihood fit to the forest at 
$N_\ion{H}{i} = 10^{13.64-16} \ \mathrm{cm}^{-2}$ from
Fig.~\ref{fig_dndz1}.}
\label{fig_dndz2}
\end{figure}

Fig.~\ref{fig_dndz2} is similar to Fig.~\ref{fig_dndz1},
except for the $N_\ion{H}{i}$ range: $N_\ion{H}{i} = 10^{13.1-14}
\ \mathrm{cm}^{-2}$. 
The correction for incompleteness due to line blending
is still negligible in this column density range
(Fig.~\ref{fig_ddf} shows that the number of lines per unit
column density over $N_\ion{H}{i} = 10^{13.1-14} \ \mathrm{cm}^{-2}$
is still well represented by a single power-law).
Again, the square from Penton et al. (\cite{pen00}) is estimated
from the equivalent widths with the assumed $b=25$ km s$^{-1}$.
The dot-dashed line is the maximum-likelihood fit to the lower column
density forest of the UVES and the HIRES data: 
$dn/dz = (55.91 \pm 2.00) \, (1+z)^{1.10 \pm
0.21}$. At $2.4 < z  < 4$ and at $2.1 < z < 4$,
$\gamma =  0.90 \pm 0.29$ and $\gamma = 1.00 \pm 0.22$, respectively.
For the column density range $N_\ion{H}{i} = 10^{13.1-14} \
\mathrm{cm}^{-2}$, the forest does not show any strong evolution.

Note that the point at $<\!z\!> \ = 2.66$ (diamond) 
from Kirkman \& Tytler (\cite{kir97}) indicates 
a number density twice as large as than at $<\!z\!> \ = 2.87$
in the interval $N_\ion{H}{i} = 10^{13.1-14.0} \ \mathrm{cm}^{-2}$
(excluding the $<\!z\!> \ = 2.66$ forest, the maximum-likelihood
fit becomes $dn/dz = (47.77 \pm 1.84) \, (1+z)^{1.18 \pm
0.22}$).
Although this discrepancy could result from a real cosmic variance of
the number density from sightline to sightline, 
the number density in the interval $N_\ion{H}{i} = 10^{13.64-16.0} 
\ \mathrm{cm}^{-2}$ from the same line of sight is in good agreement with
other HIRES data. The differential
density distribution function (Fig.~\ref{fig_ddf}) and the mean \ion{H}{i}
opacity (Fig.~\ref{fig_tau}) towards this line of sight suggest that
the discrepancy at $N_\ion{H}{i} = 10^{13.1-14.0} \ \mathrm{cm}^{-2}$
is due to overfitting, which, as discussed in Sect. 3, may occur
especially in high S/N data.

As previously noticed (Kim et al. \cite{kim97}),
the lower column density forest evolves at a slower rate 
than the higher
column density forest. The evolutionary rate $\propto (1+z)^{1.10 \pm 0.21}$
would be consistent with no evolution for $q_\mathrm{o} = 1$ or
mild evolution for $q_\mathrm{o} = 0.5$.
For $\Omega_\Lambda = 0.7$, $\Omega_\mathrm{m} = 0.3$ and
$\Omega = 1$, the Ly$\alpha$ forest with
$N_\ion{H}{i} = 10^{13.1-14} \ \mathrm{cm}^{-2}$ is mildly evolving
at $z > 1.5$.
The Ly$\alpha$ forest with
$N_\ion{H}{i} = 10^{13.1-14} \ \mathrm{cm}^{-2}$ appears 
more numerous at $z \sim 0$ than
expected when extrapolating from the $z > 1.5$ range.

\subsection{The lower cutoff $b$ in the $N_\ion{H}{i}$--$b$ distribution}

\subsubsection{The lower cutoff $b$ parameter}

For a photoionized gas, 
a temperature-density relation exists, i.e. the equation of state:
$T=T_\mathrm{0} \, (1+\delta)^{\gamma_{T}-1}$,
where $T$ is the gas temperature, $T_{0}$ is the gas temperature
at the mean gas density, $\delta$ is the baryon overdensity,
$(\rho_{b}-\overline{\rho}_{b})/\overline{\rho}_{b}$ 
($\overline{\rho}_{b}$ is the mean baryon density), and $\gamma_{T}$
is a constant which depends on the reionization history
(Hui \& Gnedin \cite{hui97a}). 
For an abrupt reionization at $z \gg 5$, the temperature of the mean gas
density decreases as $z$ decreases after the reionization,
eventually approaching an asymptotic $T_\mathrm{0}$. 
For a generally assumed
QSO-dominated UV background with a sudden turn-on of QSOs at $5 < z < 10$, 
$T_{0}$ decreases as $z$ decreases at $2 < z < 4$
(Hui \& Gnedin \cite{hui97a}).

Under the assumption that there are
some lines which are broadened primarily by the thermal motion
at any given column density, this equation of state translates
into a lower cutoff $b(N_\ion{H}{i})$
envelope in the $N_\ion{H}{i}$--$b$ distribution: $T$ and
$\delta$ can be derived from $b$ and $N_\ion{H}{i}$.
For the equation of state
$T=T_{0}\,(1+\delta)^{\gamma_{T}-1}$,
$b_{c}(N_\ion{H}{i})$ becomes
\begin{equation}
\log(b_{c}) = \log(b_{0}) + (\Gamma_{T}-1) \,
\log(N_\ion{H}{i}),
\label{eq0}
\end{equation}
where $\log(b_{0})$ is the intercept of the cutoff 
in the $\log (N_\ion{H}{i})$-$\log b$ diagram and $(\Gamma_{T}-1)$
is the slope of the cutoff
(Schaye et al. \cite{sch99}).
The cutoff slope $(\Gamma_{T}-1)$ is proportional
to $(\gamma_{T}-1)$.
This cutoff envelope provides a probe of the gas temperature of the IGM
at a given $z$, thus giving a powerful constraint on the thermal
history of the IGM (Hu et al. \cite{hu95};
Lu et al. \cite{lu96}; Kim et al. \cite{kim97};
Kirkman \& Tytler \cite{kir97}; Zhang et al. \cite{zha97};
Schaye et al. \cite{sch99};
Bryan \& Machacek \cite{bry00}; McDonald et al. \cite{mc00};
Ricotti et al. \cite{ric00};
Schaye et al. \cite{sch00}).

In practice, defining $b_{c} (N_\ion{H}{i})$
in an objective manner is not trivial due
to the finite number of available absorption lines,
sightline-to-sightline
cosmic variances, limited S/N,
and unidentified metal lines.
Among several methods proposed to derive $b_{c}
(N_\ion{H}{i})$, we have adopted the following three:
the iterative power-law fit, the power-law fit
to the smoothed $b$ distribution, and the $b$ distribution.
We refer the reader to other papers for more methods
to derive $b_{c} (N_\ion{H}{i})$ (Hu et al. \cite{hu95};
McDonald et al. \cite{mc00};
Theuns \& Zaroubi \cite{the00}).

In our analysis, we divide the data points into 2 groups:
Sample A and Sample B.
Sample A consists of the lines in the range
$N_\ion{H}{i} = 10^{12.5 -14.5} \ \mathrm{cm}^{-2}$ 
with errors less than 25\% in both $N_\ion{H}{i}$ and $b$ 
in order to avoid ill-fitted values from VPFIT.
Sample B consists of all the lines with 
$N_\ion{H}{i} = 10^{12.5 -14.5} \ \mathrm{cm}^{-2}$ regardless of errors.
The criteria for Sample A and Sample B are chosen to compare our results with 
the previous results
by Schaye et al. (\cite{sch00}) and to investigate whether it is
reasonable to include the \object{Q0302--003} line list 
for which
error estimates are not given. As no errors are 
available for \object{Q0302--003},
no Sample A can be defined at $<\!z\!> \ = 2.87$.
Note that including
the relatively few lines 
with $N_\ion{H}{i} = 10^{14.5 -16} \ \mathrm{cm}^{-2}$ 
does not change
the results significantly.
There is hardly any overlap in $z$, except for
\object{J2233-606} and \object{HE2217-2818}. Since one of our
aims is to probe the $z$-evolution
of $b_{c} (N_\ion{H}{i})$, we analyze the $b$ distribution
of each line of sight individually to derive $b_{c} (N_\ion{H}{i})$.

\begin{figure*}
\resizebox{\hsize}{!}{\includegraphics{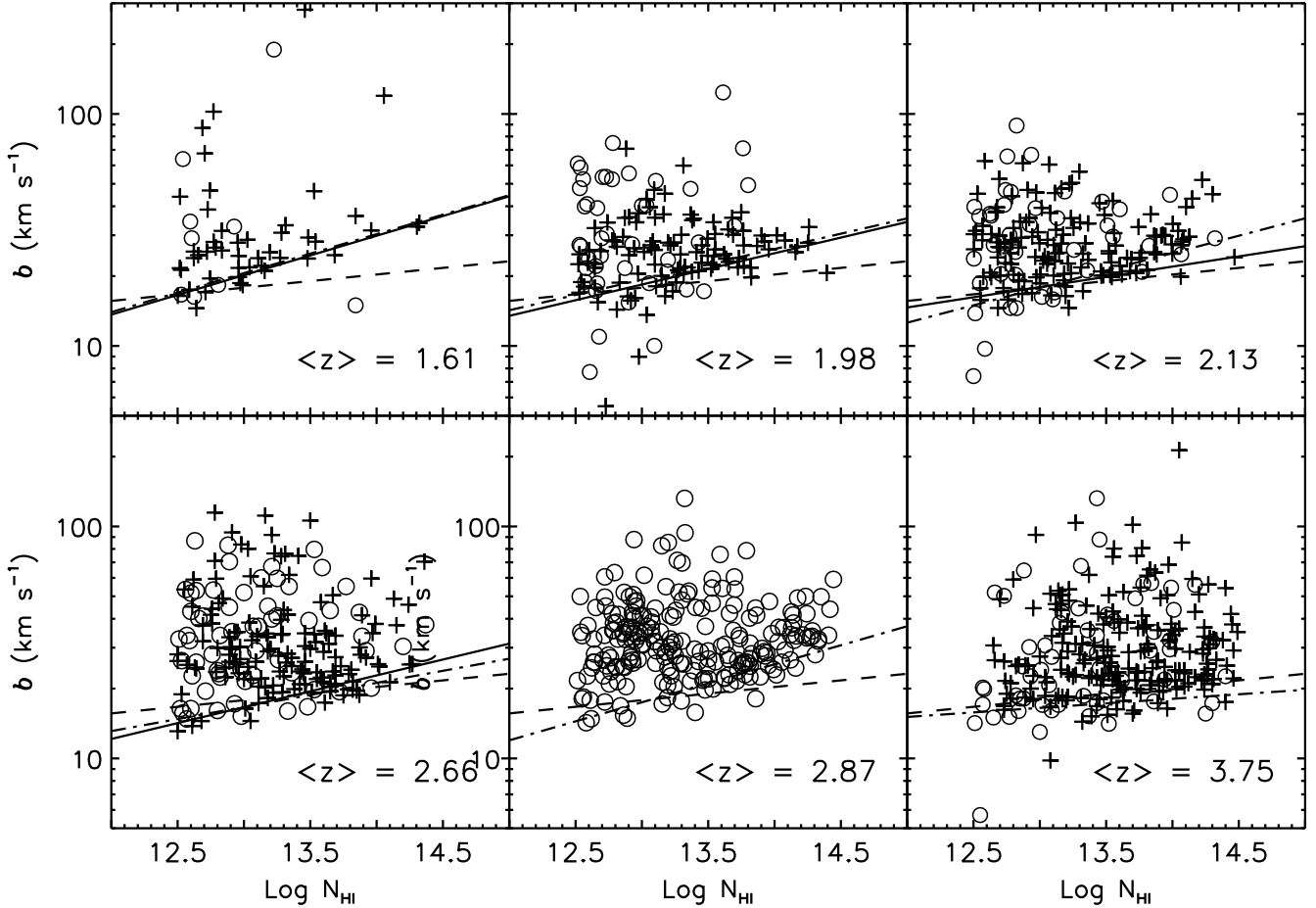}}
\caption{The $\log N_\ion{H}{i}$--$b$ diagrams with the power law
fits over $N_\ion{H}{i} = 10^{12.5-14.5} \ \mathrm{cm}^{-2}$.
Errors are not displayed.
Crosses indicate the data points with 
errors less than 25\% both
in $N_\ion{H}{i}$ and $b$ (Sample A), while
open circles indicate the lines with errors greater than
25 \% (Sample B consists of crosses and open circles together).
At $<\!z\!> \ = 2.87$, no Sample A can be defined 
due to the lack of errors in the line list. Thus, Sample B 
was used as a substitute for Sample A when compared to Sample A
from QSOs at other $z$.
The solid line, the dot-dashed line and the dashed line at each
panel indicate
the power law fit to Sample A, to Sample B and to the $<\!z\!> \ = 3.75$
forest for comparisons, respectively. At $<\!z\!> \ = 2.87$
the solid line is not present since no Sample A can be defined.} 
\label{fig_blogn}
\end{figure*}

\subsubsection{The iterative power law fit}

Since the equation of state is a power law, it is reasonable
to fit $N_\ion{H}{i}$--$b_{c}$ with one.
We did so, using the bootstrap method described by Schaye et al. (\cite{sch99}),
iterating until convergence was reached. After each iteration,
those points were excluded that lay more than one 
mean absolute deviation above the fit.
Finally, the lines more
than one mean absolute deviation below the fit were also taken out
and the final power law fit, $b_{c} = c_{0,p} \, 
N_\ion{H}{i}^{\Gamma_{T}-1}$,
was carried out.
The procedure was repeated over 200 bootstrap realizations in order to
determine the full probability distribution of the parameters of
the cutoff.
As noted by Schaye et al. (\cite{sch00}), the power law fit requires
over 200 available lines to reach stable fit parameters.

Fig.~\ref{fig_blogn} shows the iterative power law fit in the
$\log N_\ion{H}{i}$--$b$ distributions. 
The noticeable difference between Sample A (cross symbols) 
and Sample B (cross symbols and open circles)
occurs at $N_\ion{H}{i} \le 10^{13} \ \mathrm{cm}^{-2}$. These lines
usually come from blends or from weak, asymmetric absorption lines.
Table~\ref{Tab4} lists the fitted parameters, such as $c_{0,p}$
and $(\Gamma_{T}-1)$,
including $b_{c}$ values at the fixed column density $N_\ion{H}{i} = 10^{13.5} 
\ \mathrm{cm}^{-2}$, $b_{c, 13.5}$, for Sample A and Sample B.

The power law fit between Sample A
and Sample B does not give a significant difference except at
$<\!z\!> \ = 2.13$ and at $<\!z\!> \ = 3.75$, for which 
several lines with $b \le 20$ km s$^{-1}$
and $N_\ion{H}{i} \le 10^{13} \ \mathrm{cm}^{-2}$ contribute to a 
different power law fit for Sample B. 
This suggests that using the \object{Q0302--003} line
list at $<\!z\!> \ = 2.87$ without error bars does not severely 
distort our conclusions. Note that
the power law fit at $<\!z\!> \ = 2.87$ might be less steep
with a higher intercept,
if the same general behavior of errors also occurs for 
the \object{Q0302--003} forest (larger errors at
$b \le 20$ km s$^{-1}$ or $b \ge 40$ km s$^{-1}$).
Due to the small number of lines (47 lines for
Sample A and 56 lines for Sample B) at $<\!z\!> \ = 1.61$, 
the power law fit 
should be taken as an upper limit on $b_{c} (N_\ion{H}{i})$ and indeed it
provides the highest $b_{c, 13.5}$ among all the $z$ bins.
For both Sample A and Sample B, there is a weak trend of
increasing $b_{c, 13.5}$ as $z$ decreases, except at $<\!z\!> \ = 2.87$
which shows a higher $b_{c, 13.5}$ value than at the adjacent $z$ ranges
(see Sect. 6.2 for further discussion). 
On the other hand, the power law slope $(\Gamma_{T}-1)$ 
is rather ill-defined with $z$ with a possible
flatter slope at $<\!z\!> \ = 3.75$ than at $z < 3.1$.

\begin{figure}
\resizebox{\hsize}{!}{\includegraphics{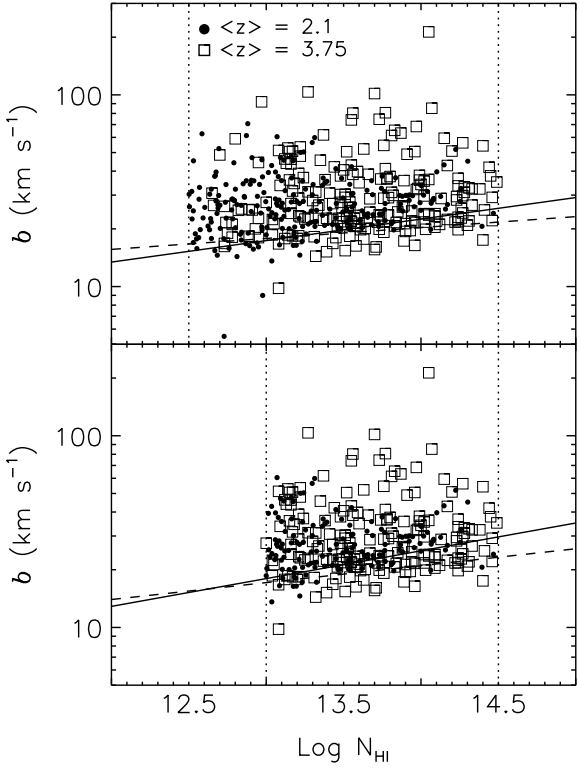}}
\caption{The $\log N_\ion{H}{i}$--$b$ diagrams with the power law
fits at $z \sim 2.1$ (only for \object{J2233--606} and
\object{HE2217--2818}; small filled circles) and at 
$<\!z\!> \ =3.75$ (open squares).
At both redshifts, the data points with errors less than 25\% in both 
$N_\ion{H}{i}$
and $b$ are displayed.
The upper panel shows the fit to the column density range
$N_\ion{H}{i} = 10^{12.5-14.5} \ \mathrm{rm}^{-2}$, while
the lower panel is to the range $N_\ion{H}{i} = 10^{13-14.5} \ 
\mathrm{rm}^{-2}$.
The vertical dotted lines in each panel represent the column density
range over which the power law fit was carried out.
The solid line shows the power law fit at $z \sim 2.1$, and the
dashed line at $z=3.75$.
A deficit of 
lines with $N_\ion{H}{i} \le 10^{13} \
\mathrm{cm}^{-2}$ and $b \le 20$ km s$^{-1}$
observed at $<\!z\!> \ = 3.75$ is due in part to the fact
that all the fitted
lines with $N_\ion{H}{i} \le 10^{13} \
\mathrm{cm}^{-2}$ and $b \le 20$ km s$^{-1}$ have errors larger than
25 \% and in part to the severe line blending which 
limits the detection of weak lines.}
\label{fig_test1}
\end{figure}

Fig.~\ref{fig_test1} shows the power law fit to Sample A
at $z \sim 2.1$ (small filled circles; 
242 lines from \object{J2233--606} and
\object{HE2217--28118})
and at $<\!z\!> \ = 3.75$ 
(open squares; 209 lines) over $N_\ion{H}{i} = 10^{12.5-14.5} \
\mathrm{cm}^{-2}$ (upper panel) and over
$N_\ion{H}{i} = 10^{13-14.5} \
\mathrm{cm}^{-2}$ (lower panel). 
The fitted parameters are given in Table~\ref{Tab4}.
For both $N_\ion{H}{i}$ ranges, 
the slopes of $b_{c} (N_\ion{H}{i})$ are steeper
at $z \sim 2.1$ than at $<\!z\!> \ = 3.75$.
This result, however, is certainly biased by the lack of
lines with $b \le 15$ km s$^{-1}$ and 
$N_\ion{H}{i} \le 10^{13.4} \mathrm{cm}^{-2}$ at higher $z$,
due to the severe line blending.

\begin{table*}
\caption[]{The power law fit to the $N_\ion{H}{i}$--$b$ distributions}
\label{Tab4}
\begin{tabular}{crccccrccc}
\hline
\noalign{\smallskip}
& \multicolumn{4}{c}{Sample A} &
& \multicolumn{4}{c}{Sample B} \\
\\[-2ex]
\cline{2-5} \cline{7-10} \\[-2ex]
$<\!z\!>$ & \#$^{\mathrm{a}}$ & $\log (c_{0,p})$ & 
$(\Gamma_{T}-1)$ & $b_{c, 13.5}$ (km
s$^{-1}$)$^{\mathrm{b}}$ & 
& \#$^{\mathrm{a}}$ & $\log (c_{0,p})$ & 
$(\Gamma_{T}-1)$ & $b_{c, 13.5}$ (km
s$^{-1}$)$^{\mathrm{b}}$ \\
\noalign{\smallskip}
\hline
\noalign{\smallskip}
1.61 & 47 & $-0.92 \pm 0.09$ & $0.17 \pm 0.01$ & $24.5 \pm 2.4$ & 
& 56 & $-0.87 \pm 0.10$ & $0.17 \pm 0.01$ & $24.9 \pm 1.1$ \\
1.98 & 103 & $-0.49 \pm 0.11$ & $0.14 \pm 0.01$ & $21.4 \pm 0.5$ & 
& 146 & $-0.43 \pm 0.12$ & $0.13 \pm 0.01$ & $21.4 \pm 0.6$ \\
2.13 & 139 & $0.11 \pm 0.08$ & $0.09 \pm 0.01$ & $20.1 \pm 0.6$ & 
& 181 & $-0.70 \pm 0.11$ & $0.15 \pm 0.01$ & $20.6 \pm 0.5$ \\
2.66 & 140 & $-0.55 \pm 0.10$ & $0.14 \pm 0.01$ & $19.6 \pm 1.1$ & 
& 204 & $-0.13 \pm 0.10$ & $0.10 \pm 0.01$ & $18.7 \pm 0.5$ \\
2.87 & - & - & - & - & & 223 & $-0.89 \pm 0.08$ & $0.16 \pm 0.01$ 
& $20.4 \pm 0.8$ \\
3.75 & 209 & $0.51 \pm 0.06$ & $0.06 \pm 0.01$ & $18.7 \pm 1.2$ & 
& 271 & $0.71 \pm 0.10$ & $0.04 \pm 0.01$ & $17.1 \pm 0.9$ \\
2.1$^{\mathrm{c}}$ & 242 & $-0.22 \pm 0.08$ & $0.11 \pm 0.01$ & 
$19.6 \pm 0.6$ &
& 327 & $-0.73 \pm 0.08$ & $0.15 \pm 0.01$ & $20.4 \pm 0.5$ \\ 
2.1$^{\mathrm{d}}$ & 156 & $-0.64 \pm 0.13$ & $0.15 \pm 0.01$ & 
$21.2 \pm 0.7$ &
& 187 & $-0.78 \pm 0.12$ & $0.16 \pm 0.01$ & $20.9 \pm 0.6$  \\ 
3.75$^{\mathrm{e}}$ & 188 & $0.09 \pm 0.10$ & $0.09 \pm 0.01$ & 
$18.9 \pm 1.0$ &
& 233 & $-0.06 \pm 0.09$ & $0.10 \pm 0.01$ & $18.1 \pm 0.9$ \\ 
\noalign{\smallskip}
\hline
\end{tabular}
\begin{list}{}{}
\item[$^{\mathrm{a}}$] The number of absorption lines used for the fit.
\item[$^{\mathrm{b}}$] The errors were estimated from the difference
between the mean $b_{c, 13.5}$ and the minimum/maximum $b_{c, 13.5}$
from the 200 bootstrap realizations for the given $N_\ion{H}{i}$--$b$
pairs. The typical $1\sigma$ of the $b_{c, 13.5}$ distribution
from the 200 bootstrap realizations is $\sim 0.3$ km sec$^{-1}$, which is
underestimated (cf. Schaye et al. \cite{sch00}).
\item[$^{\mathrm{c}}$] Combined line lists from 
\object{J2233--606} and \object{HE2218-2817}.
The power law fit was carried out for 
$N_\ion{H}{i} = 10^{12.5-14.5} \ \mathrm{cm}^{-2}$.
\item[$^{\mathrm{d}}$] Combined line lists from 
\object{J2233--606} and \object{HE2218-2817}.
The power law fit was carried out for 
$N_\ion{H}{i} = 10^{13-14.5} \ \mathrm{cm}^{-2}$.
\item[$^{\mathrm{e}}$] For 
$N_\ion{H}{i} = 10^{13-14.5} \ \mathrm{cm}^{-2}$.
\end{list}
\end{table*}

\subsubsection{The power law fit to the smoothed $b$ distribution} 

Bryan \& Machacek (\cite{bry00}) presented a method to measure 
$b_{c} (N_\ion{H}{i})$ from a power law fit to 
a smoothed $b$ distribution, sorting absorption lines
by $N_\ion{H}{i}$ and then dividing them into groups
containing similar numbers of lines.
The $b$ distribution in each group was then smoothed with a Gaussian filter with
a smoothing constant $\sigma_{b}$: 
\begin{equation}
S_{b,j} (b) = \sum_{i} \exp(-(b_{i}-b)^{2}/2\sigma_{b}^{2}),
\end{equation}
where $S_{b,j} (b)$ is the smoothed density of lines in each group $j$
and $i$ indicates the lines in the group.
Then, the location of the first peak in the derivative of $S_{b,j} (b)$ 
defines the lower cutoff
at the average column density, $N_{\ion{H}{i},j}$, 
for the $j$-th group. 

\begin{figure}
\resizebox{\hsize}{!}{\includegraphics{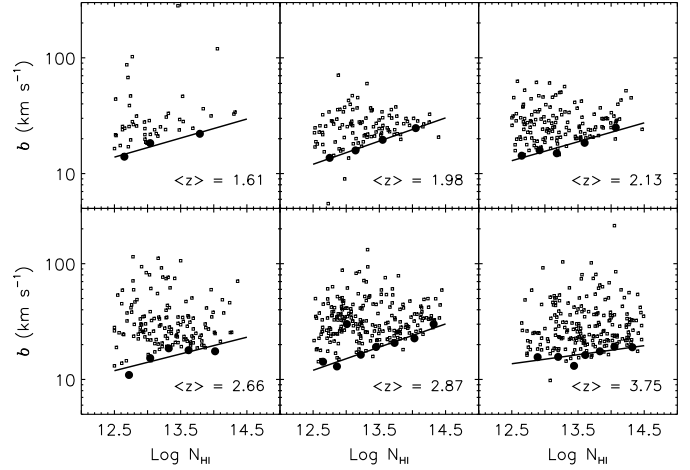}}
\caption{The $\log N_\ion{H}{i}$--$b$ diagrams for Sample A at
each $z$ (Sample B at $<\!z\!> \ = 2.87$)
with the robust least-squares power law fit to the
smoothed $b$ distribution 
with the Gaussian smoothing constant 3 km s$^{-1}$.
Small open squares represent the data points for Sample A
(Sample B at $<\!z\!> \ = 2.87$), while
filled circles represent the cutoff $b$ values estimated from
the smoothed $b$ distributions with $\sim 30$ lines in each group 
(see the text for the details). 
The solid line represents the robust least-squares power law fit
to filled circles.}
\label{fig_bryan}
\end{figure}

Fig.~\ref{fig_bryan} shows the $\log N_\ion{H}{i}$--$b$ diagram at each 
$z$ with the
$b_{c} (N_\ion{H}{i})$ points for each group (filled circles) 
measured from the smoothed 
$b$ distributions. 
We use the smoothing constant $\sigma_{b} = 3$ km s$^{-1}$. However,
$S_{b,j} (b)$ is largely insensitive to the smoothing constant.
In general, 30 lines were included in each group except
for the last group at higher $N_\ion{H}{i}$ for which typically smaller
numbers of lines
were available.
For this same reason, at $<\!z\!> \ = 1.61$
groups of $\sim 16$ lines were used.
The solid line represents the 
robust least-squares power law fit
to filled circles: $b_{c} (N_\ion{H}{i}) 
= c_{0,s} N_\ion{H}{i}^{\ \Gamma_{T}-1}$.
Table~\ref{Tab5} lists the parameters of 
the power law fit to the smoothed $b$
distributions.

We find that the power law fit to the smoothed $b$ distribution
produces in general a lower intercept and a steeper slope 
than the iterative power law fit. It also produces smaller
$b_{c, \mathrm{13.5}}$ values.
Direct comparison of Fig.~\ref{fig_bryan} with Fig.~\ref{fig_blogn}
indicates that $b_{c} (N_\ion{H}{i})$ measured 
from the smoothed $b$ distribution
can be considered as a lower limit on the {\it real}
$b_{c} (N_\ion{H}{i})$, while $b_{c} (N_\ion{H}{i})$ from the
iterative power law fit can be considered as an upper limit
on the {\it real} $b_{c} (N_\ion{H}{i})$.

As with the iterative power law fit, 
$b_{c, \mathrm{13.5}}$ measured from the smoothed $b$ distribution
increases continuously as $z$ decreases,
except at $<\!z\!> \ = 2.87$, where
$b_{c, \mathrm{13.5}}$ is higher than at the adjacent redshifts
(see Sect. 6.2 for further discussion).
The slope $(\Gamma_{T}-1)$ measured from the smoothed $b$ distribution 
also does not show any well-defined trend with
$z$. 

\begin{table}
\caption[]{The power law fit to the smoothed $b$ distributions}
\label{Tab5}
\begin{tabular}{cccc}
\hline
\noalign{\smallskip}
$<\!z\!>$ & $\log(c_{0,s})$ & 
$(\Gamma_{T}-1)$ & 
$b_{c, 13.5}$ \\
& & & (km s$^{-1}$) \\
\noalign{\smallskip}
\hline
\noalign{\smallskip}
1.61 & $-0.92 \pm 0.13$ & $0.16 \pm 0.04$ & $20.3 \pm 1.1$ \\
1.98 & $-1.42 \pm 0.02$ & $0.20 \pm 0.01$ & $19.1 \pm 1.0$ \\
2.13 & $-0.93 \pm 0.11$ & $0.16 \pm 0.03$ & $18.8 \pm 1.1$ \\
2.66 & $-0.73 \pm 0.22$ & $0.14 \pm 0.06$ & $16.6 \pm 1.2$ \\
2.87$^{\mathrm{a}}$ & $-1.42 \pm 0.08$ & $0.20 \pm 0.02$ & 
$19.0 \pm 1.1$ \\
3.75 & $0.16 \pm 0.07$ & $0.08 \pm 0.02$ & $16.4 \pm 1.1$ \\
2.1$^{\mathrm{b}}$ & $-0.71 \pm 0.14$ & $0.15 \pm 0.04$ & $19.2 
\pm 1.0$ \\
2.1$^{\mathrm{c}}$ & $-0.50 \pm 0.34$ & $0.13 \pm 0.09$ & 
$19.7 \pm 1.1$ \\
3.75$^{\mathrm{d}}$ & $-0.30 \pm 0.13$ & $0.11 \pm 0.03$ & 
$15.8 \pm 1.0$ \\
\noalign{\smallskip}
\hline
\end{tabular}
\begin{list}{}{}
\item[$^{\mathrm{a}}$] For Sample B since Sample A cannot be defined.  
\item[$^{\mathrm{b}}$] At $N_\ion{H}{i} = 10^{12.5-14.5} \ \mathrm{cm}^{-2}$
for \object{J2233--606} and \object{HE2218-2817}.
\item[$^{\mathrm{c}}$] At $N_\ion{H}{i} = 10^{13-14.5} \ \mathrm{cm}^{-2}$
for \object{J2233--606} and \object{HE2218-2817}.
\item[$^{\mathrm{d}}$] At $N_\ion{H}{i} = 10^{13-14.5} \ \mathrm{cm}^{-2}$.
\end{list}
\end{table}

\subsubsection{The $b$ distributions}

Assuming that absorption lines arise from
local optical depth ($\tau$) peaks and that $\ln \tau$ is a Gaussian random
variable, Hui \& Rutledge (\cite{hui99}) derived a single-parameter $b$ 
distribution:
\begin{equation}
dn/db = B_\mathrm{HR} \, \frac{{b_\sigma}^4}{b^5}\,
\exp(-\frac{{b_\sigma}^4}{b^4}),
\end{equation}
where $n$ is the number of absorption lines, 
$B_\mathrm{HR}$ is a 
constant and $b_\mathrm{\sigma}$
is a parameter determined by the average amplitude of the fluctuations and the
effective smoothing scale.

\begin{figure}
\resizebox{\hsize}{!}{\includegraphics{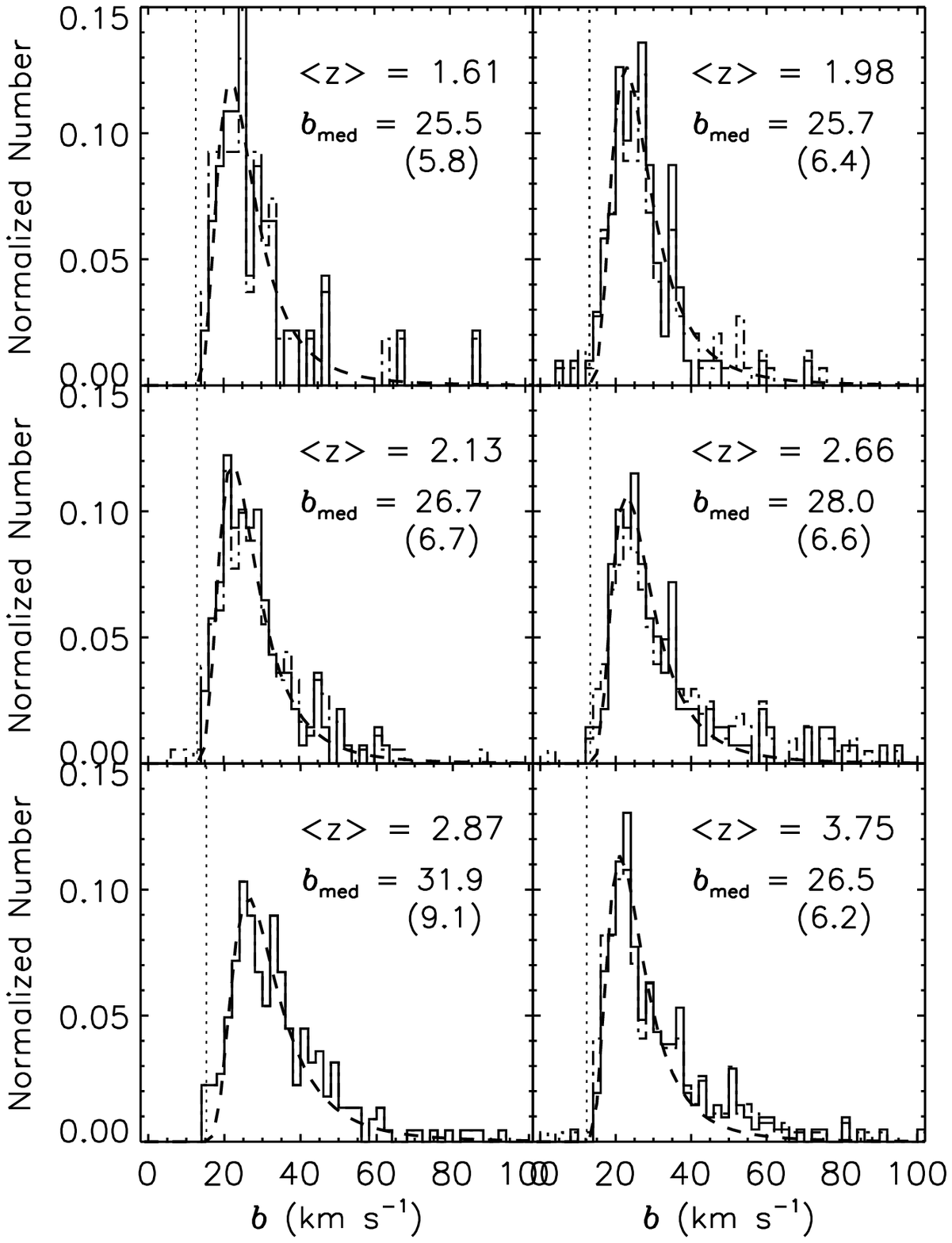}}
\caption{The $b$ distribution of the Ly$\alpha$ forest at each $z$. 
While solid lines are for
Sample A, dot-dashed lines are for Sample B
(no Sample A at $<\!z\!> \ = 3.75$). 
The dashed line and the dotted line represent
the best-fitting Hui-Rutledge function and $b_\mathrm{HR}$,
respectively.
The $b_\mathrm{med}$ value in the panels indicates the
median $b$ value at the corresponding $z$ for Sample A.
The number in parentheses indicates the $1\sigma(b)$ value from
the Gaussian distribution.}
\label{fig_bdis}
\end{figure}

Fig.~\ref{fig_bdis} shows the observed $b$ distributions
at each $z$. The
noticeable difference between Sample A (solid lines) and 
Sample B (dot-dashed lines)
occurs at $b \le 20$ km s$^{-1}$ or $b \ge 40$ km s$^{-1}$.
These unphysical lines are usually introduced by VPFIT  
to fit noises so that the overall profile of \ion{H}{i} forest complexes
could be improved.
The dashed line represents
the best-fitting Hui--Rutledge $b$ distribution, while
the dotted line represents the $b$ parameter for which 
the Hui-Rutledge $b$ distribution function vanishes
to $10^{-4}$, $b_\mathrm{HR}$, i.e. 
the truncated $b$ value for the Hui-Rutledge $b$
distribution function.
The parameter $b_\mathrm{HR}$ cannot be considered equivalent to 
the cutoff $b_{c}$
since it is derived from the $b$ distribution
without assuming the $b_{c}$ dependence on $N_\ion{H}{i}$.
It is more sensitive to smaller $b$ values in the $b$ distribution,
which are in general coupled with lower $N_\ion{H}{i}$.
Table~\ref{Tab6} lists the relevant parameters describing 
the Hui-Rutledge $b$ distribution for Sample A,
such as the constant $B_\mathrm{HR}$, $b_\sigma$, $b_\mathrm{HR}$ and the
median $b$ values at different column density ranges. 

It is hard to specify subtle differences among the $b$
distributions: while the modal $b$ value and the $b_\mathrm{HR}$ value
have a 
minimum at $<\!z\!> \ =3.75$, they have a maximum at $<\!z\!> \ =2.87$.
The $<\!z\!> \ =2.87$ forest
also has the broadest $b$ distribution. However, this large $\sigma(b)$
could be in part due to a different fitting program
and in part due to a lack of information on the errors.
Other parameters, such as
$b_\sigma$, $b_\mathrm{HR}$, and $\sigma (b)$, appear to be fairly constant
with $z$.

\begin{figure}
\resizebox{\hsize}{!}{\includegraphics{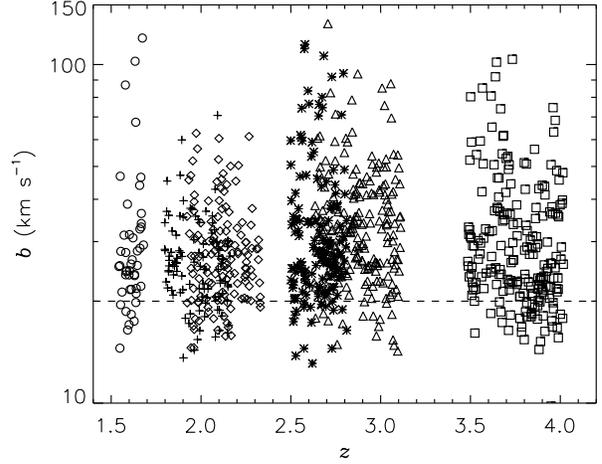}}
\caption{The $b$ distribution as a function of $z$ for Sample A
(Sample B at $<\!z\!> \ = 3.75$).
The horizontal
dashed line indicates $b = 20$ km s$^{-1}$ which is a 
$N_\ion{H}{i}$-independent $b_{c}$
at $z \sim 2.8$ (Hu et al. \cite{hu95}).
Circles, crosses, diamonds, stars, 
triangles and squares are from \object{HE0515--4414},
\object{J2233--606}, \object{HE2217--2818}, \object{HS1946+7658},
\object{Q0302--003} and \object{Q0000--263}, respectively.
There is an indication of increasing $b_{c}$ with decreasing
$z$ at $z \sim 3.7$. At $z < 3.1$, $b_{c}$ is not clearly
defined.}
\label{fig_bz}
\end{figure}

Fig.~\ref{fig_bz}
shows the $b$ distribution with $z$. This diagram 
does not assume a $b_{c}$ dependence on $N_\ion{H}{i}$,
but is sensitive to a local $b_{c} (N_\ion{H}{i})$ variance.
At $z < 3.1$,
there is no clear indication 
of the behavior of the lower cutoff $b$ values as a function of $z$. 
However, there is
a clear indication of a trend with $z$ of the lower cutoff $b$ 
values over $3.5 < z < 3.9$.
In Fig.~\ref{fig_bz}, there are distinct regions at $1.8 < z < 2.4$.
The apparent cutoff values in $b$ at $2.2 < z < 2.4$ and at $1.8 < z < 1.9$
are clearly higher than at $1.9 < z < 2.2$. The
$2.2 < z < 2.4$ region towards \object{HE2217--2818}
corresponds to a $\sim 44 \, h^{-1}$ Mpc void (the region
B in Fig.~\ref{fig_he22void}), which might suggest enhanced
ionization due to a nearby QSO or processes of 
galaxy formation (Theuns et al. \cite{the00a}).


\begin{table}
\caption[]{The parameters of the $b$ distributions}
\label{Tab6}
\begin{tabular}{cccccc}
\hline
\noalign{\smallskip}
$<\!z\!>$ & $B_\mathrm{HR}$ & $b_\sigma$ & $b_\mathrm{HR}$ &
$b_\mathrm{med}^{\mathrm{a}}$ & $b_\mathrm{med}^{\mathrm{b}}$ \\
& & (km s$^{-1}$) & (km s$^{-1}$) & (km s$^{-1}$) & (km s$^{-1}$) \\
\noalign{\smallskip}
\hline
\noalign{\smallskip}
1.61 & 7.37 & 23.01 & 12.61 & 28.14 & 34.56 \\
1.98 & 7.98 & 23.83 & 13.04 & 26.04 & 29.10 \\
2.13 & 7.46 & 23.61 & 12.95 & 25.34 & 29.57 \\
2.66 & 6.82 & 24.09 & 13.25 & 28.30 & 30.10 \\
2.87 & 7.09 & 27.75 & 15.30 & 28.74 & 34.05 \\
3.75 & 6.72 & 22.41 & 12.31 & 28.90 & 30.70 \\
\noalign{\smallskip}
\hline
\end{tabular}
\begin{list}{}{}
\item[$^{\mathrm{a}}$] For lines with 
$N_\ion{H}{i} = 10^{13.1-14} \ \mathrm{cm}^{-2}$ 
from Sample A (Sample B at $<\!z\!> \ = 2.87$).
\item[$^{\mathrm{b}}$] For lines with 
$N_\ion{H}{i} = 10^{13.8-16} \ \mathrm{cm}^{-2}$ 
from Sample A (Sample B at $<\!z\!> \ = 2.87$).
\end{list}
\end{table}

\subsection{The two-point velocity correlation function}

The Ly$\alpha$ forest contains information on the large-scale matter
distribution and the simplest way to study it is to compute the
two-point velocity correlation function, $\xi(\Delta v)$.
The correlation function compares the observed
number of pairs ($N_\mathrm{obs}$) with the expected number of pairs
($N_\mathrm{exp}$) from a random distribution in a given velocity bin
($\Delta v$):
$\xi(\Delta v) = N_\mathrm{obs}(\Delta v)/N_\mathrm{exp}(\Delta v)
-1$, where $\Delta v = c\,(z_{2}-z_{1})/[1+ (z_{2}+z_{1})/2]$, 
$z_{1}$ and $z_{2}$ are redshifts of two lines and $c$ is the
speed of light (Cristiani et al. \cite{cri95}; Cristiani et al. \cite{cri97};
Kim et al. \cite{kim97}).

Studies of the correlation function of the Ly$\alpha$ forest have
generally led to conflicting results even at similar $z$. Some
studies find a lack of clustering (Sargent et al. \cite{sar80} at $1.7
< z < 3.3$; Rauch et al. \cite{rau92} at $z \sim 3$; Williger et
al. \cite{wil94} at $z \sim 4$), while others find clustering at 
scales $\Delta v \le 350$ km s$^{-1}$ (Webb \cite{web87}
at $1.9 < z < 2.8$;
Hu et al. \cite{hu95} at $z \sim 2.8$; 
Kulkarni et
al. \cite{kul96} at $z \sim 1.9$; Lu et al. \cite{lu96} at $z \sim
3.7$; Cristiani et al. \cite{cri97} at $z \sim 3.3$).

\begin{figure}
\resizebox{\hsize}{!}{\includegraphics{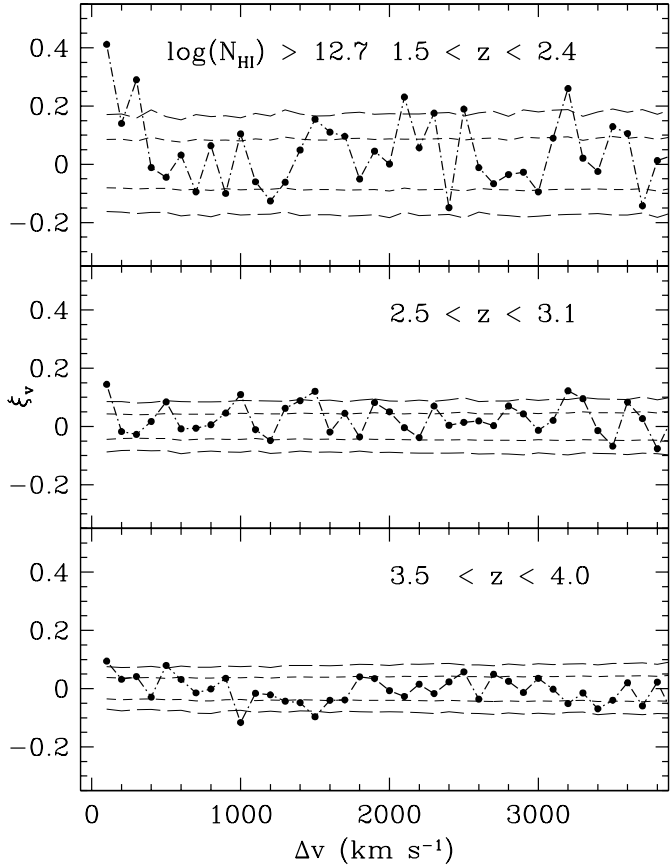}}
\caption{Evolution of the two-point correlation function with redshift 
for Ly$\alpha$ lines with column densities above
$N_\ion{H}{i} = 10^{12.7} \ \mathrm{cm}^{-2}$. Short-dashed and
long-dashed lines represent the $1 \sigma$ and $2 \sigma$ Poisson
errors.}
\label{fig_velcor}
\end{figure}

Fig.~\ref{fig_velcor} shows the velocity correlation strength
at $\Delta v < 4\,000$ km s$^{-1}$. 
To obtain sufficient statistics, the analysis was carried out in
three redshift bins: $1.5<z<2.4$ (\object{HE0515--4414},
\object{J2233--606}, and \object{HE2217--2818}), 
$2.5<z<3.1$ (\object{HS1946+7658} and \object{Q0302--003}) and 
$3.5<z<4.0$ (\object{Q0000--263}).

In our approach $N_\mathrm{exp}$ was estimated averaging $1\,000$ numerical
simulations of the observed number of lines, trying to account for
relevant cosmological and observational effects. In particular
a set of lines was randomly generated in the same redshift
interval as the data according to the cosmological distribution
$\propto (1+z)^{\gamma}$, with $\gamma = 2.4$ (see Sect. 4.2). The
results are not sensitive to the value of $\gamma$ adopted and even a
flat distribution (i.e. $\gamma =0$) gives values of $\xi$ that differ
typically by less than 0.02. Line blanketing of weak lines due to strong
complexes was also accounted for. Lines with too small velocity
splittings, compared with the finite resolution or the intrinsic
blending due to the typical line widths--the so-called
``line-blanketing'' effect (Giallongo et al. \cite{gia96}), were 
excluded in the estimates of $N_\mathrm{exp}$.

Clustering is clearly detected at low redshift: at $1.5<z<2.4$ in the
100 km s$^{-1}$ bin, we measure $\xi = 0.4\pm 0.1$ for lines with 
$N_\ion{H}{i} \ga 10^{12.7} \ \mathrm{cm}^{-2}$. 
There is a hint of increasing amplitude with
increasing column density:
in the same redshift range $\xi = 0.35\pm 0.08$ for lines with 
$N_\ion{H}{i} \ga 10^{12.5} \ \mathrm{cm}^{-2}$. 
The trend is not significant but 
agrees with the behavior observed at higher redshifts 
(Cristiani et al. \cite{cri97}; Kim et al. \cite{kim97}).
Unfortunately the number of lines observed in the interval $1.5<z<2.4$
does not allow us to extend the analysis to higher column densities,
although groups of strong lines are occasionally evident
(e.g. the range 3230--3270 \AA\/ in \object{HE0515--4414}).

The amplitude of the correlation at $100$ km s$^{-1}$ decreases
significantly with increasing redshift from $0.4 \pm 0.1$ at
$1.5<z<2.4$, to $0.14\pm0.06$ at $2.5<z<3.1$ and $0.09\pm0.07$ at
$3.5<z<4.0$.

\begin{table*}
\caption[]{Voids at $z \sim 2$}
\label{Tab7}
\begin{tabular}{lccccccc}
\hline
\noalign{\smallskip}
QSO & Region & Wavelength & $<\!z\!>$ & $\Delta z$ & Comoving size$^{\mathrm{a}}$ 
& $x_\mathrm{gap}$ & $P_{>}(x_\mathrm{gap})$ \\
& & (\AA\/) & & & ($h^{-1}$ Mpc) & & \\
\noalign{\smallskip}
\hline
\noalign{\smallskip}
\object{HE0515--4414} & A & 3088--3161 & 1.570 & 0.060 & 61.1 & 5.7 & 0.045 \\
\object{HE2217--2818} & A & 3504--3579 & 1.913 & 0.062 & 54.3 & 8.4 & 0.012 \\
\object{HE2217--2818} & B & 3878--3946 & 2.218 & 0.056 & 43.5 & 8.0 & 0.018 \\
\noalign{\smallskip}
\hline
\end{tabular}
\begin{list}{}{}
\item[$^{\mathrm{a}}$] For $q_\mathrm{0}=0.1$.
\end{list}
\end{table*} 

\subsection{Voids}

Voids along the three low-redshift lines of sight were searched for.
For comparison with previous results (Carswell \& Rees
\cite{car87}; Crotts \cite{cro87}; Ostriker et al. \cite{ost88}),
we identify a void as a region
without any absorption stronger than $N_\ion{H}{i} \sim 10^{13.5} \
\mathrm{cm}^{-2}$ over a comoving size of
at least $30\,h^{-1}$ Mpc (assuming $q_\mathrm{0}=0.1$).

\begin{figure}
\resizebox{\hsize}{!}{\includegraphics{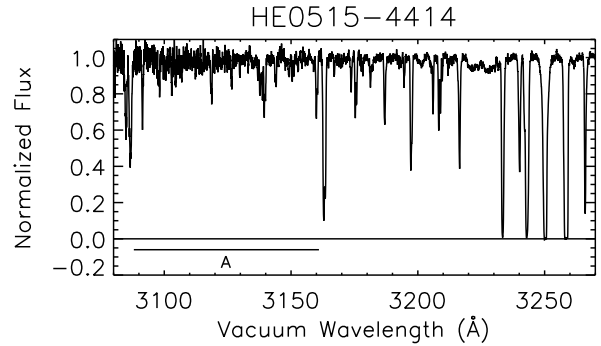}}
\caption{The spectrum of \object{HE0515--4414} with the void
at $z = 1.570$.
See the text for the details.}
\label{fig_he05void}
\end{figure}

\begin{figure}
\resizebox{\hsize}{!}{\includegraphics{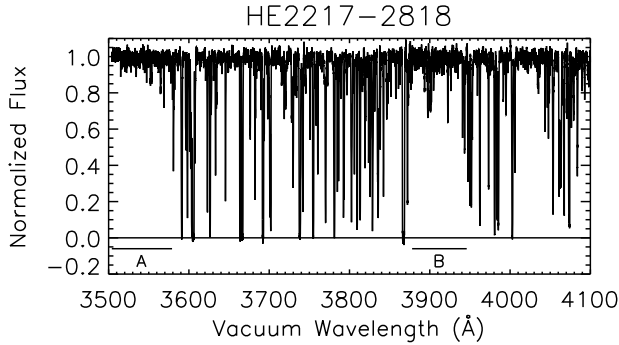}}
\caption{The spectrum of \object{HE2217--2818} with the void regions.
The voids are indicated as A at $z = 1.913$ and B at $z = 2.218$.
See the text for the details.}
\label{fig_he22void}
\end{figure}

Figs.~\ref{fig_he05void}--\ref{fig_he22void} show the voids
detected in the spectrum of \object{HE0515--4414} and
\object{HE2217--2818}, respectively. No significant void was found in
the spectrum of \object{J2233--606}. The wavelength range used for 
searching for voids has been
selected to be redward of the QSO's Ly$\beta$ emission line and
$3\,000$ km s$^{-1}$ blueward of the QSO's Ly$\alpha$ emission to
avoid the proximity effect. The wavelength range searched for
voids is larger than 
that used to study the Ly$\alpha$ forest in other sections. 

Table~\ref{Tab7} lists the dimensions of the voids,
as well as the probability of finding a void larger than their 
comoving size. The probability was calculated assuming a Poisson
distribution of the local forest. In this case, the probability 
of finding a void larger than a given size $x_\mathrm{gap}$ is
$P_{>}(x_\mathrm{gap}) = 1 - (1-\exp^{-x_\mathrm{gap}})^{n}$,
where $x_\mathrm{gap}$ is the line interval in the unit of the local
mean line interval and $n$ is the number of lines with 
$N_\ion{H}{i} \ge 10^{13.5} \ \mathrm{cm}^{-2}$
(Ostriker et al. \cite{ost88}). 
The joint probability of finding two voids with a size larger than 
$40 \, h^{-1}$ Mpc at $z \sim 2$, as observed in the spectrum of
\object{HE2217--2818}, is of the order of $2 \times 10^{-4}$.
The results correspond very well
to the probability estimates derived from the 
simulations described above.

There are different ways to produce a void in the forest: a large
fluctuation in the gas density of absorbers, enhanced UV
ionizing radiation from nearby faint QSOs or star-forming galaxies,
feedback processes (including shock heating) from 
galaxy formation
(Dobrzycki \& Bechtold \cite{dob91}; Heap et al. \cite{hea00};
Theuns et al. \cite{the00a}). 
We recall here that the void B in the spectrum of \object{HE2217--2818}
corresponds to a region of above-average Doppler parameter (Sect.~4.3.3).
It will be interesting to carry out deep imaging around
\object{HE2217--2818} to identify QSO candidates and investigate
whether a local ionizing source is responsible for the $\sim
50\,h^{-1}$ Mpc voids.

\section{The flux statistics on the Ly$\alpha$ forest}

The traditional Voigt profile fitting analysis
is limited by two major drawbacks. 
First, there is no unique solution.
Although the $\chi^{2}$ minimization is normally applied to the fit,
different fitting programs produce slightly different results.
Even using the same program, different $\chi^{2}$ thresholds
lead to different numbers of lines when line blending is severe.
As the resolution and S/N increase,
many forest lines show various degrees of departure from the
Voigt profile. This departure can be fitted by adding 
physically improbable narrow components
in high S/N data to improve
the overall fit,
while the same profile can usually be
fitted by one broad component in low S/N data (Schaye et
al. \cite{sch99}; Theuns \& Zaroubi \cite{the00}). Thus, it becomes
harder to compare fitted line parameters from different observations
and different fitting programs, as the resolution and S/N increase.

Second, there is no guarantee that absorption lines have 
a Voigt profile. Although searches for
characteristic non-Voigt signatures have been inconclusive
(Rauch \cite{rau96}; Outram et al. \cite{out99b}), 
numerical simulations predict
non-Voigt profiles of absorption lines
(Outram et al. \cite{out00}).
In numerical simulations, 
lines are broadened by a combination of gas temperature, peculiar
velocity, Hubble expansion,
and Jeans smoothing (Bryan et al. \cite{bry99};
Theuns \& Zaroubi \cite{the00}).
In conclusion, the profile fitting method should be regarded as
a powerful parameterization of the spectrum, not a representation
of the physics governing forest clouds
(Rauch et al. \cite{rau92}; Hu et al. \cite{hu95};
Kirkman \& Tytler \cite{kir97}; Bryan et al. \cite{bry99};
Schaye et al. \cite{sch99};
Theuns \& Zaroubi \cite{the00}).

In order to avoid the non-uniqueness of profile fitting 
analyses and to allow
more straightforward comparisons with theoretical predictions,
a direct use of observed spectra of the Ly$\alpha$
forest has been explored. 
The most straightforward way to characterize the
observed
spectra is to use the N-point functions of the transmitted flux, $F$, 
or the observed
optical
depth, $\tau$ 
(Miralda-Escud\'e et al. \cite{mir97}; Rauch et al. \cite{rau97};
Zhang et al. \cite{zha98}; Bryan et al. \cite{bry99}; 
Machacek et al. \cite{mac00};
Theuns et al. \cite{the00b}). In the following, we
consider 
the one-point function and the two-point function
of the transmitted flux as well as other statistical 
measures such as the line
count and the optical depth correlation function.

For comparison with the forest at $z > 2.4$, we generated artificial
spectra from the published line lists. These lists are the same
as those used in the Voigt profile fitting analysis in Sect. 4 
(see Table~\ref{Tab2}). We added Gaussian noise
to the artificial spectra, according
to the quoted S/N. However, as Theuns et al. (\cite{the00b}) stated,
Gaussian noise independent of flux 
does not represent the real, observed S/N.
The difference becomes more evident at 
$F \sim 0$ (saturated regions), where Gaussian noise produces
larger fluctuations than what is observed.
Unfortunately, without details on the number of individual
spectra for a given wavelength range, data reduction and the
normalization of the spectra, the noise at the bottom of saturated lines
cannot be simulated correctly just from the published S/N and the
given CCD readout noise.

We tested our approach with two extreme cases: artificial
spectra without noise and artificial spectra with Gaussian noise.
Except for $F \sim 0$ at $<\!z\!> \ = 3.75$, 
the results from these two cases do not 
differ significantly, in particular, at $0.1 < F < 0.8$ where most
results are considered. We use the results from the spectra generated
with Gaussian noise in this study. However,
we mention any difference between these two
extreme cases when it becomes noticeable. Also
note that most simulations do not include noise in their analysis and, in 
a sense noise-free spectra would be more appropriate 
to be compared with the results from simulations.
Due to the imperfect simulation of the noise in the artificially 
generated spectra,
the results drawn from the flux statistics 
should be taken qualitatively at $F \sim 0$ and $F \sim 1$.

In order to be consistent with our profile fitting analysis,
we excluded high-column-density systems with 
$N_\ion{H}{i} > 10^{16} \ \mathrm{cm}^{-2}$
from the spectra. Note that including these high-column-density systems
does not change the general qualitative conclusions from the flux statistics.
It only changes the {\it quantitative} results considerably at $z \sim 3.7$ 
when $F$ becomes close to 0 due to the higher number of pixels
with $F \sim 0$ at $z \sim 3.7$.
However, for the estimation of the mean \ion{H}{i} opacity, 
we used the whole 
regions including high-column-density absorption systems 
in order to compare with the previous results
from the literature (Press et al. \cite{pre93}; Rauch et al. \cite{rau97}).
We also remind the reader to be particularly cautious in using 
the line list of \object{Q0302--003} at $z \sim 2.9$, 
which is the only one {\it not}
generated by VPFIT. As for the Voigt fitting analysis, a
systematic difference in $N_\ion{H}{i}$ and $b$ could change
the results in this Sect. at $z \sim 2.9$.

\subsection{The mean \ion{H}{i} opacity}

The \ion{H}{i} opacity, $\tau_\ion{H}{i} (\lambda)$,
is defined as $\tau_\ion{H}{i} (\lambda) = - \ln
(F_{\lambda}/F_\mathrm{c})$,
where $\lambda$ is the observed wavelength, $F_{\lambda}$
is the observed flux at $\lambda$, and $F_\mathrm{c}$ is
the continuum flux at
$\lambda$. Since the opacity
scales logarithmically, the mean opacity cannot be measured
accurately when $F_{\lambda} \approx 0$. The
effective opacity, $\tau_\mathrm{eff}$, is typically used in place of
$\overline{\tau}_\ion{H}{i}$: $\exp^{-\tau_\mathrm{eff}} =\
<\!\exp^{-\tau}\!>$, where
$<\,\,>$ indicates the mean value
averaged over wavelength. Note that the mean \ion{H}{i}\ opacity
and the effective \ion{H}{i}\
opacity are different quantities. However, we refer to
the estimated $\tau_\mathrm{eff}$ values as 
the ``mean \ion{H}{i}\ opacity" in this study.

\begin{figure}
\resizebox{\hsize}{!}{\includegraphics{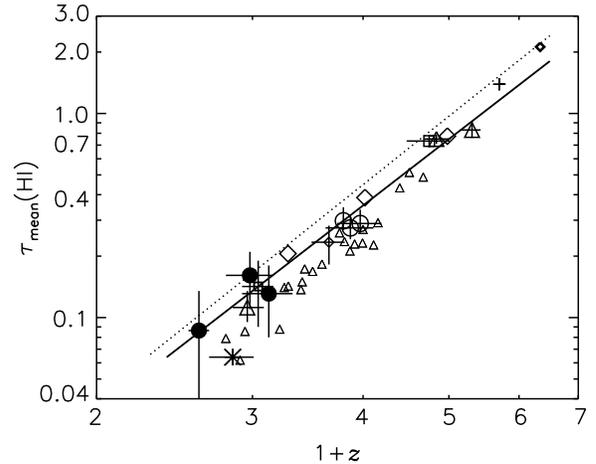}}
\caption{The \ion{H}{i} opacity as a function of $(1+z)$. Filled circles
represent the mean \ion{H}{i}\ opacity from the UVES data.
Small triangles and diamonds represent
$\overline{\tau}_\ion{H}{i}$ from
Zuo \& Lu (\cite{zuo93}; adding up the
equivalent widths from the published line lists) and Rauch et al.
(\cite{rau97}; we used their $\overline{\tau}_\ion{H}{i}^{\mathrm{cent}}$
uncorrected for continuum fitting uncertainty
since all other observations were
not corrected for continuum fitting uncertainties),
respectively. Open circles, the star, the large square, 
the small diamond
and the small square are estimated from
the spectra generated artificially using the line list provided by
Hu et al. (\cite{hu95}), Kulkarni et al. (\cite{kul96}),
Lu et al. (\cite{lu96}), Kirkman \& Tytler (\cite{kir97}), and
Outram et al. (\cite{out99a}), respectively.
Large triangles were read from Schaye et al. (\cite{sch00}).
All the x-axis error bars represent the $z$ range used for the
$\overline{\tau}_\ion{H}{i}$ estimates, while the y-axis error bars
were estimated from simply changing the adopted continuum by $\pm 5$\%,
except for large triangles whose y-axis errors were read from
Schaye et al. (\cite{sch00}). The cross at $z=4.7$
(Songaila et al. \cite{son99}) and
the thick diamond at $z=5.34$ (Spinrad et al. \cite{spi98})
are from the Keck\thinspace II/LRIS data.
The dotted line represents
the commonly used formula by Press et al. (\cite{pre93}):
$\overline{\tau}_\ion{H}{i} (z) =
0.0037 \, (1+z)^{3.46}$, given by the
conventional power law fit. The solid line represents
the power law fit to the UVES and the HIRES data.}
\label{fig_tau}
\end{figure}

Fig.~\ref{fig_tau} shows our $\overline{\tau}_\ion{H}{i}$
measures (in fact, the effective
optical depth $\tau_\mathrm{eff}$), together with
other opacity measures compiled from the literature.
Filled circles are the $\overline{\tau}_\ion{H}{i}$ measures
from the UVES data including high-column-density absorption systems.
Note that all the other opacities from the literature 
(except the Press et al. (\cite{pre93}) $\overline{\tau}_\ion{H}{i}$
measurements)
were measured including
high-column-density regions in the spectra except damped Ly$\alpha$
systems. The Press et al. measure (dotted line)
was derived 
at low resolution, 
including damped Ly$\alpha$ systems.
Our experiments with the UVES data
show that there is no noticeable
difference between the $\overline{\tau}_\ion{H}{i}$ measured from
the spectra generated from the line lists
and the $\overline{\tau}_\ion{H}{i}$ from the observed spectra.
This indicates that very weak lines do not significantly
contribute to $\overline{\tau}_\ion{H}{i}$.
Therefore, the estimated $\overline{\tau}_\ion{H}{i}$ 
from the spectra generated from the published line lists 
using high resolution, high S/N data can be considered to be
reliable. In Table~\ref{Tab8}, we list the estimated
$\overline{\tau}_\ion{H}{i}$ values when these values are not given in
{\it numeric form} in the relevant references of Fig.~\ref{fig_tau}.

\begin{table*}[t]
\caption[]{The mean \ion{H}{i} opacity}
\label{Tab8}
\begin{tabular}{lcccc}
\hline
\noalign{\smallskip}
QSO & $<\!z\!>$ & $z_\mathrm{Ly\alpha}$ & $\tau_\ion{H}{i}$ & Ref. \\
\noalign{\smallskip}
\hline
\object{HE0515--4414} & 1.61 & 1.54--1.68 & 0.086$^{0.049}_{-0.051}$ & This
study \\
\object{Q1331+170} & 1.85 & 1.68--2.01 & 0.064$^{0.049}_{-0.051}$ &
Kulkarni et al. (\cite{kul96}) \\
\object{Q1100--264} & 1.96 & 1.85--2.09 & 0.112$^{0.023}_{0.017}$
& Schaye et al. (\cite{sch00}) \\
\object{J2233--606} & 1.98 & 1.80--2.17 & 0.161$^{0.049}_{-0.051}$ &
This study \\
\object{J2233--606} & 2.04 & 1.92--2.17 & 0.142$^{0.049}_{-0.051}$ &
Outram et al. (\cite{out99a}) \\
\object{HE2217--2818} & 2.13 & 1.92--2.33 & 0.131$^{0.049}_{-0.051}$
& This study \\
\object{HS1946+7658} & 2.66 & 2.50--2.81 & 0.234$^{0.049}_{-0.051}$
& Kirkman \& Tytler (\cite{kir97}), HIRES \\
\object{Q0636+680} & 2.80 & 2.58--3.02 & 0.298$^{0.049}_{-0.051}$ &
Hu et al. (\cite{hu95}), HIRES \\
\object{Q0302--003} & 2.87 & 2.63--3.11 & 0.275$^{0.049}_{-0.051}$ &
Hu et al. (\cite{hu95}), HIRES \\
\object{Q0014+813} & 2.97 & 2.74--3.20 & 0.289$^{0.049}_{-0.051}$ &
Hu et al. (\cite{hu95}), HIRES \\
\object{Q0000--263} & 3.75 & 3.48--4.02 & 0.733$^{0.049}_{-0.051}$
& Lu et al. (\cite{lu96}), HIRES \\
\object{Q2237--061} & 3.84 & 3.69--4.02 &
0.75$^{0.04}_{-0.04}$ & Schaye et al. (\cite{sch00}), HIRES \\
\object{Q2237--061} & 4.31 & 4.15--4.43 & 0.83$^{0.06}_{-0.08}$ &
Schaye et al. (\cite{sch00}), HIRES \\
\noalign{\smallskip}
\hline
\end{tabular}
\end{table*}

The opacity measures are dependent on the continuum fitting.
For low-resolution data, the continuum is usually extrapolated
from longward of the QSO's Ly$\alpha$ emission line, typically resulting
in an overestimation of the continuum, i.e. an overestimated
$\overline{\tau}_\ion{H}{i}$. On the other hand, the local continuum
fitting generally adopted for high-resolution data may result in an
underestimation of the continuum, i.e. an underestimated
$\overline{\tau}_\ion{H}{i}$.
Therefore, it is not surprising that
the $\overline{\tau}_\ion{H}{i}$
(dotted line) from low-resolution data by Press et al. (\cite{pre93})
is higher than any other measurements. Small triangles from Zuo \&
Lu (\cite{zuo93}) were estimated from the published
line lists using intermediate resolution data, which usually 
do not include low column density lines.
Thus, the Zuo \& Lu estimates
are about a factor of 2 lower
than the PRS formula at $z \sim 3$. Other observations
fall inbetween the PRS formula and the Zuo \& Lu
values.

There is a scatter in $\overline{\tau}_\ion{H}{i}$ at similar $z$,
even though opacities are estimated from data
with the same instrument configuration and reduction, 
such as two filled circles at $z \sim 2.1$
from UVES and three open circles at $z \sim 2.8$ from
HIRES. This scatter could result from measurement
errors, such as the continuum fitting, and from a cosmic spatial
variance. The continuum fitting becomes unreliable at $z > 3$
due to severe line blending.
On the other hand, the cosmic variance is present
at all $z$.
For example, the difference
between two UVES measurements at $z \sim 2.1$ (filled circles)
is due to the two high-column-density systems with
$N_\ion{H}{i} \ge 10^{17} \ \mathrm{cm}^{-2}$ towards
\object{J2233--606}.
When these high-column-density systems are excluded,
the opacity at $<\!z\!>\,=1.98$ (\object{J2233--606})
becomes 0.123, which is similar to the one at $<\!z\!>\,=2.13$
(\object{HE2217--2818}). Similarly, the higher opacity at
$z \sim 2.80$ towards \object{Q0636+680} is due to several
high-column-density clouds
on this line of sight, while two other lines of sight at
$z \sim 2.8$ are relatively devoid of high-column-density
clouds.
Despite the same HIRES configurations,
the $\overline{\tau}_\ion{H}{i}$ values from Hu et al. (\cite{hu95})
and from Kirkman \& Tytler (\cite{kir97}) are a factor of 1.2
lower than the Rauch et al. values. 
Since the line lists for the Rauch et al. QSO sample are not
published,
we cannot test whether the presence of high-column-density systems
in their sample causes the higher $\overline{\tau}_\ion{H}{i}$.

The solid line represents the least-squares
fit to the UVES and the HIRES data:
$\overline{\tau}_\ion{H}{i} (z) =
0.0030 \pm 0.0008 \, (1+z)^{3.43\pm 0.17}$.
The new UVES data at $1.5 < z < 2.4$ suggest that
$\overline{\tau}_\ion{H}{i}$ can be well approximated by a single power law
at $1.5 < z < 4$. Two $\overline{\tau}_\ion{H}{i}$ measures
at $z > 4.5$ (cross and thick diamond)
from the Keck\thinspace II/LRIS data 
suggest that $\overline{\tau}_\ion{H}{i}$
might be significantly higher than extrapolated from
$z < 4$.
However, these values were derived from low-resolution data, which
usually overestimate $\overline{\tau}_\ion{H}{i}$. In fact,
they correspond better to the Press et al. formula, which
was also derived from low-resolution data. Without more
high-resolution data at
higher $z$, it is premature to conclude that the $\overline{\tau}_\ion{H}{i}$
evolution at $z > 4$ departs significantly from a single power law (cf.
Schaye et al. \cite{sch00}).

In the standard Friedmann universe with the cosmological
constant $\Lambda =0$, if the baryon overdensity
$\delta$ is $\delta \le 10$, the mean \ion{H}{i} opacity can
be expressed by
$\overline{\tau}_\ion{H}{i} \propto (1+z)^{3.3}/ \Gamma_\ion{H}{i} (z)$
(Machacek et al. \cite{mac00}).
Although there are a few high-column-density systems
in Fig.~\ref{fig_tau},
our result on
$\overline{\tau}_\ion{H}{i}$, $\overline{\tau}_\ion{H}{i}
\propto (1+z)^{3.43 \pm 0.17}$ at $1.5 < z < 4$,
is in good agreement with the predicted power law index. This suggests
that
$\Gamma_\ion{H}{i} (z)$ does not strongly evolve over this $z$ range.

\subsection{The one-point function of the flux}

The one-point function of the flux (or the probability
density distribution function of the transmitted flux), $P (F)$,
is simply the number of 
pixels which have a flux between $F$ and $F + dF$ for a given flux $F$
over the entire number of pixels per $dF$.
In other words, it is the probability density to find a pixel at
a given $F$
(Miralda-Escud\'e et al. \cite{mir97}; 
Rauch et al. \cite{rau97}; 
Bryan et al. \cite{bry99}; Machacek et al. \cite{mac00}; 
Theuns et al. \cite{the00b}).

\begin{figure}
\resizebox{\hsize}{!}{\includegraphics{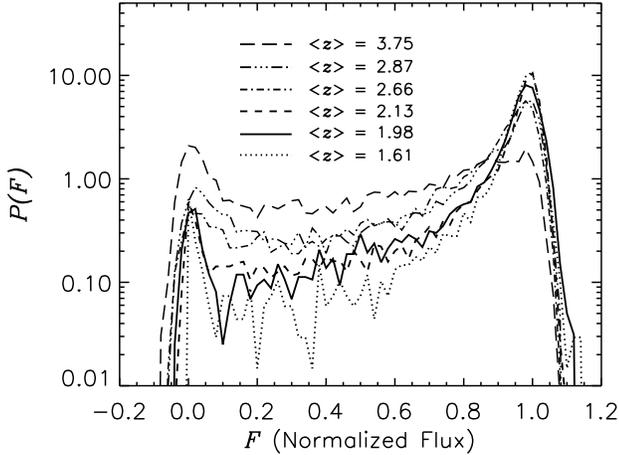}}
\caption{The one-point function of the flux, $P (F)$,
as a function of $F$ at different $z$. Strong absorption 
lines (smaller $F$) disappear more rapidly than 
weak absorption lines (larger $F$) as $z$ decreases.}
\label{fig_fdd}
\end{figure}

Fig.~\ref{fig_fdd} shows $P (F)$ as a function of $F$.
The one-point function of the flux at $F < 0$ and $F > 1$
from observations is due 
to observational and continuum fitting uncertainties.
The non-smooth $P (F)$ at $<\!z\!> \ = 1.61$ is due to the small
number of pixels used to calculate $P (F)$.
The wider $P (F)$ profiles at $F \sim 0$ and at $z > 2.4$
compared with at $z < 2.4$ are
due to the characteristics of Gaussian noise in the
spectra generated (Theuns et al. \cite{the00b}). 
For the spectra generated without noise, the $P (F)$ profiles at 
$z > 2.4$ are narrower with higher amplitude at $F \sim 0$,
but do not differ significantly from the $P(F)$ profiles in
Fig.~\ref{fig_fdd} at $0.2 < F < 0.8$.

The flattening towards $F \sim 1$ at
$<\!z\!> \ = 3.75$ is due to 
the smaller number of $F \sim 1$ pixels from
the increasing forest number density at higher $z$.
The one-point functions of the flux at $z \sim 2.1$
(for \object{J2233--606} and \object{HE2217--2818})
are a factor of 1.3 higher than the sCDM model 
simulated by Machacek et al. (\cite{mac00})
at $0.2 < F < 0.6$ (this lower simulated
$P(F)$ is in agreement with their lower mean \ion{H}{i} 
opacity at $z \sim 2$).

After smoothing $P (F)$ at $<\!z\!> \ = 1.61$ over a $dF = 0.14$
bin, $P(F)$ at $F=0.2$ becomes $P(F)_\mathrm{F=0.2} \propto \,
(1+z)^{3.86 \pm 0.54}$. The flux $F=0.2$ corresponds 
to $N_\ion{H}{i} \sim 3.2 \times 10^{13} \ \mathrm{cm}^{-2}$
(in this study, we assume $b=30$ km s$^{-1}$).  
At $F=0.6$ ($N_\ion{H}{i} = 1.0 \times 10^{13} \ \mathrm{cm}^{-2}$),
$P(F)_\mathrm{F=0.6} \propto \, 
(1+z)^{3.26 \pm 0.21}$.
In short, the probability density of 
finding strong absorption lines (smaller $F$)
shows a steeper slope than that of 
finding weak absorption lines (larger $F$). This 
is in good agreement with the result from the Voigt profile fitting
analysis: the higher column density forest disappears more rapidly than
the lower column density forest as $z$ decreases.

\subsection{The two-point function of the flux}

The two-point function of the flux, $P(F_{1},F_{2},\Delta v)$, is
the probability of two pixels with the $\Delta v$ velocity separation having
normalized fluxes $F_{1}$ and $F_{2}$. It is usually expressed as
\begin{eqnarray*}
\overline{\Delta F} (\Delta v,\delta F_{1})
\end{eqnarray*}
\begin{equation}
\equiv \int_{\delta F_{1}}\, \left[\int^{\infty}_{-\infty}
(F_{1}-F_{2})P(F_{1},F_{2},\Delta v)\,dF_{2}\right] \, dF_{1}/\delta F_{1},
\end{equation}
where $\overline{\Delta F} (\Delta v,\delta F_{1})$ is the mean flux difference
between two pixels with $F_{1}$ and $F_{2}$, which are separated
by $\Delta v$. At $\Delta v > 200$ km s$^{-1}$, 
$\overline{\Delta F} (\Delta v,\delta F_{1})$
is just the difference between the mean transmitted flux and 
$F_{1}$. At $30 < \Delta v < 100$ km s$^{-1}$,
the shape of $\overline{\Delta F} (\Delta v,\delta F_{1})$ contains
information on the profile shape of absorption lines. 
However, great care has to be exercised in the
interpretation
since $\overline{\Delta F} (\Delta v,\delta F_{1})$ is
an average quantity (Miralda-Escud\'e et al.
\cite{mir97}; Bryan et al. \cite{bry99}; Machacek et al.
\cite{mac00}; Theuns et al. \cite{the00b}).

\begin{figure}
\resizebox{\hsize}{!}{\includegraphics{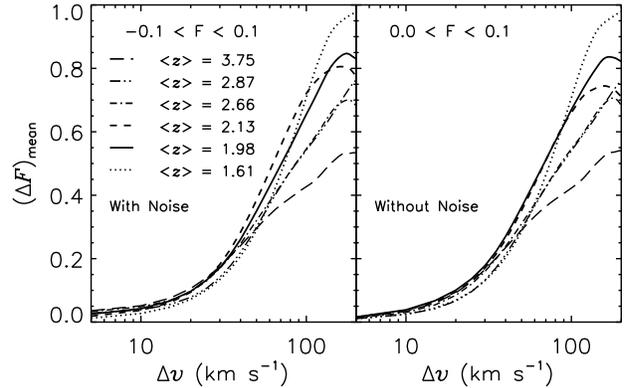}}
\caption{The mean flux difference, 
$\overline{\Delta F} (\Delta v,\delta F_{1})$,
as a function of $\Delta v$ at different $z$. 
Since $\overline{\Delta F} (\Delta v,\delta F_{1})$ is
an averaged quantity, it is not obvious to interpret 
$\overline{\Delta F} (\Delta v,\delta F_{1})$ to be a strong indicator
of the temperature of the IGM.}
\label{fig_fdd1}
\end{figure}

Fig.~\ref{fig_fdd1} shows
$\overline{\Delta F} (\Delta v,\delta F_{1})$
as a function of $\Delta v$ at different $z$. The left-hand
panel shows the results for the observed spectra 
at $z < 2.4$ and the spectra generated
with noise at $z > 2.4$, while the right-hand panel 
shows the results for
the spectra generated from the line lists for all $z$ without noise.
There is no noticeable difference between the two panels.
To be comparable
with the simulations by Machacek et al. (\cite{mac00}),
the flux range
was chosen to be $-0.1 \le \! F \! \le 0.1$, which
corresponds to $N_\ion{H}{i} \ge 10^{13.77} \ \mathrm{cm}^{-2}$.
Over this column density range, absorption lines are in general
saturated and belong to a \ion{H}{i} complex, 
rather than being isolated.

The overall $z$-dependence on $\overline{\Delta F} (\Delta v,\delta
F_{1})$ is not clear. At $\Delta v < 30$ km s$^{-1}$,
$\overline{\Delta F} (\Delta v,\delta F_{1})$ 
shows almost identical profiles at all $z$ except at 
$<\!z\!> \ =$ 1.61 and 2.87. At $<\!z\!> \ =$ 1.61 and 2.87,
$\overline{\Delta F} (\Delta v,\delta F_{1})$ is wider than at any other $z$.
At $\Delta v > 30$ km s$^{-1}$, there is a tendency for 
the width of $\overline{\Delta F} (\Delta v,\delta F_{1})$ to become
narrower as $z$ decreases.
Following Machacek et al. (\cite{mac00}),
we measured $\Delta v (0.3)$, 
the width of $\overline{\Delta F} (\Delta v,\delta F_{1})$
at which $\overline{\Delta F} (\Delta v,\delta F_{1})$
becomes 0.3. At $<\!z\!> \ =$ 1.61, 1.98, 2.13, 2.66, 2.87, and 3.75,
$\Delta v (0.3)$ is 51.06 (51.61), 44.25 (43.15), 
41.62 (44.27), 47.05 (48.54), 49.99 (50.79)
and 50.95 (52.11) km s$^{-1}$
(the number in parentheses is for the spectrum generated
without noise), respectively.
The observed $\Delta v (0.3)$ values are larger than
the predicted $\Delta v (0.3)$ values, $\sim 35$ km s$^{-1}$, 
from the models considered by Machacek et al. (\cite{mac00}) at $z \sim 3$.

A weak trend of decreasing $\Delta v (0.3)$ with decreasing $z$
might suggest that the gas temperature decreases as $z$ decreases.
In fact, Theuns et al. (\cite{the00b})
note that at a given $z$ 
a simulation with a hotter gas temperature shows
a wider $\overline{\Delta F} (\Delta v,\delta F_{1})$ profile
than a simulation with a lower gas temperature.
However, numerical simulations also show that higher $b$
values at higher $z$ may be a result of other physical 
processes. Variations of $\overline{\Delta F} (\Delta v,\delta F_{1})$
as a function of $z$ can be a result of increasing
line blending at higher $z$ as well as a change in the gas temperature.

\subsection{Line counts of the Ly$\alpha$ forest}

The line count at a given flux $F_{t}$ is defined as the number of regions
below $F_{t}$
(Miralda-Escud\'e et al. \cite{mir96}). The number defined
this way is more straightforward to determine than the conventional
line counting from the profile fitting. 

\begin{figure}
\resizebox{\hsize}{!}{\includegraphics{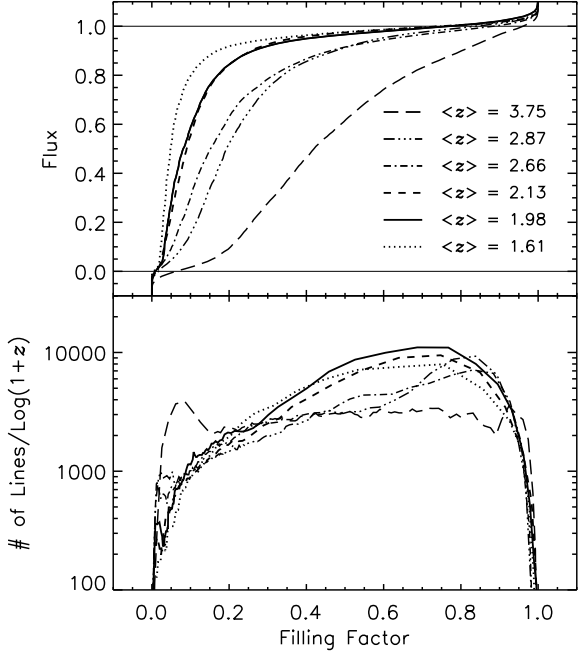}}
\vspace{-1.cm}
\caption{The line counts of the Ly$\alpha$ forest.
The upper panel shows the normalized flux as a function of 
the filling factor. 
The lower panel represents the line counts from the observed spectra
and the spectra generated with noise,
which were smoothed with a 20 km s$^{-1}$ box-car function in order to decrease
noise.}
\label{fig_linecount}
\end{figure}

The upper panel of Fig.~\ref{fig_linecount} 
shows the normalized flux  
of the observed spectra and the spectra generated 
with noise as a function of the filling factor. The filling factor
is the fraction of the spectrum occupied by 
the pixels whose normalized flux is smaller than a given $F_{t}$. 
Since this method is sensitive to noise,
the spectra were smoothed with a 20 km s$^{-1}$ box-car 
function. Except at $<\!z\!> \ = 3.75$ (the broad
feature at the filling factor close to 0
is due to the characteristics
of Gaussian noise, i.e. larger root-mean-square 
fluctuations than the real,
observed fluctuations at $F \sim 0$), the line counts are 
similar when the filling factor is 0.07--0.3.
This range of the filling factor corresponds to $0.2 < F < 0.9$
at $z \sim 1.8$ and to $0.05 < F < 0.8$ at $z \sim 2.7$, i.e.
pixels with a wide range of fluxes are considered. 
However, the same filling factor range only probes
$0 < F < 0.3$ at $<\!z\!> \ = 3.75$, where only strong lines
are counted. This results in the different line count
at $<\!z\!> \ = 3.75$.
When the filling factor is greater than 0.3, 
the line counts 
deviate from each other.
This regime corresponds to 
$F \sim 1$, where noise distorts the true
line counts.

\begin{figure}
\resizebox{\hsize}{!}{\includegraphics{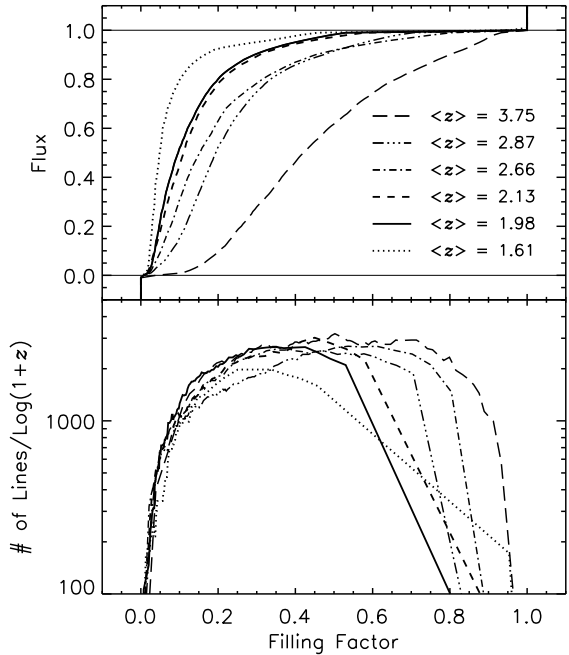}}
\vspace{-1.cm}
\caption{The line counts of the Ly$\alpha$ forest.
The upper panel shows the normalized flux as a function of
the filling factor.
The lower panel represents the line counts of the spectra 
generated artificially from the line lists.
The abrupt 
drops in the line counts at the larger filling factor 
are due to the fact that the large
number of the $F=1$ pixels do not contribute to the line counts
as the filling factor approaches 1.}
\label{fig_mkcount}
\end{figure}

Fig.~\ref{fig_mkcount} shows the line counts 
as a function of the filling factor again. In this diagram,
the line counts were calculated using the artificial spectra
generated from the fitted line lists for each QSO without
adding noise, i.e. they have an infinite S/N.
Note that this process does not include weak lines, usually not 
present in the fitted line lists.
Unlike the lower panel of Fig.~\ref{fig_linecount},
the curves describing the line counts as 
a function of the filling factor show a similar 
progression
as a function of $z$ when a filling factor is smaller
than $\sim 0.3$, 
except for $<\!z\!> \ = 2.87$. The $<\!z\!> \ = 2.87$
forest also shows
a slightly different behavior in the filling factor-flux diagram.
The flux at a given filling factor increases continuously as
$z$ decreases except that the flux corresponding to a given filling factor 
is larger at $<\!z\!> \ = 2.87$
than at $<\!z\!> \ = 2.66$ for a filling factor larger than 0.5.
This might indicate the real cosmic variance in the structure
of the Ly$\alpha$ forest along the line of
sight towards \object{Q0302--003} (one known void
towards \object{Q0302--003} at $z \sim 3.17$ is not included in this study.
Also note that the $<\!z\!> \ = 2.13$ forest includes
one void region).
However, a similar work done by Kim (\cite{kim99}) did not
show any difference in the line counts as a function of $z$
at $2.1 < z < 4$ when the real observed spectra including 
\object{Q0302--003} were used. It is highly
unlikely that the QSO sample in Kim's work shows 
the same amount of systematic
differences from the QSOs used in this study for all $z$. 
Therefore, we cast doubts on the cosmic variance as a probable 
reason for the different line counts at
$<\!z\!> \ = 2.87$. Rather, the line parameters of
\object{Q0302--003} have not been obtained with the program VPFIT
and
this suggests that a different behavior of the
Ly$\alpha$ forest at $z \sim 2.9$ from the rest of the forest
at different $z$, such as a higher $b_{c, 13.5}$ than at
adjacent $z$, 
should be taken with caution.

Since the filling factor
is determined mainly by the Hubble expansion, the negligible
$z$-dependence of the line counts suggests that the evolution of
the forest at $1.5 < z < 4$ is driven mainly by the Hubble expansion
(Miralda-Escud\'e et al. \cite{mir96}).

\subsection{The step optical depth correlation function}

Miralda-Escud\'e et al. (\cite{mir96}) first introduced
a correlation function using a pixel-by-pixel transmitted flux,
which is more straightforward than the two-point
velocity correlation function.
Cen et al. (\cite{cen98}) developed this concept
further. We analyzed the clustering properties of the Ly$\alpha$
forest, following Cen et al.'s methods (\cite{cen98}).
Among their newly defined correlation functions,
we only consider 
the step optical depth correlation.
In general, the trends we found from the step optical depth 
correlation function
hold for the other correlation functions.

The step optical depth correlation function $\xi_{\tau, s}$
is defined as
\begin{equation}
\xi_{\tau, s} (\Delta v) \equiv \frac{<\tau_{s}(v+\Delta v)
\tau_{s}(v)>}{<\tau_{s}>^{2}} -1,
\end{equation}

\noindent where the step optical depth is defined as
\begin{equation}
\tau_{s} (\Delta v) \equiv \left\{\begin{array}
{r@{ , \quad}l}
1 & \mathrm{if}\,\, \tau_\mathrm{obs}(v) \ge \tau_\mathrm{min}\, , \\
0 & \mathrm{if}\,\, \tau_\mathrm{obs}(v) < \tau_\mathrm{min}\, .
\end{array}
\right.
\end{equation}

\begin{figure}
\resizebox{\hsize}{!}{\includegraphics{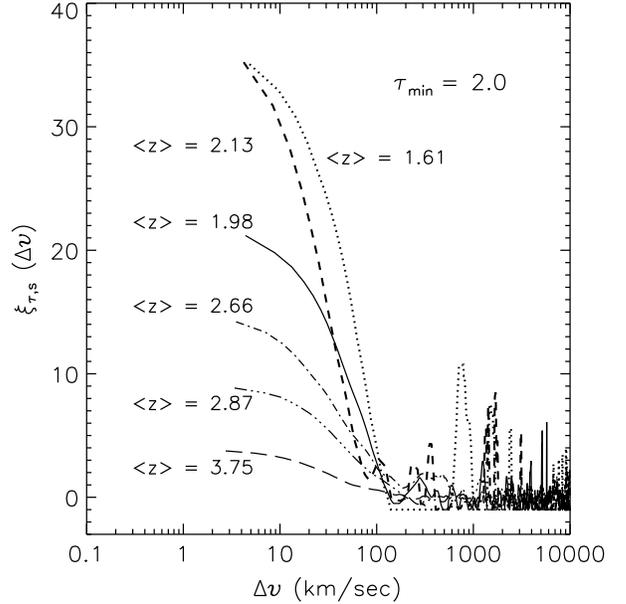}}
\caption{The step optical depth correlation functions.
They are sensitive to the profile shapes of the forest at the
smaller velocity separations for a given
$\tau_\mathrm{min}$.}
\label{fig_cor}
\end{figure}

Fig.~\ref{fig_cor} shows the step optical depth correlation functions
$\xi_{\tau, s} (\Delta v)$ with $\tau_\mathrm{min} = 2$ 
as a function of $\Delta v$. The $\tau_\mathrm{min} = 2$
corresponds to $N_\ion{H}{i} \sim 10^{13.6} \ \mathrm{cm}^{-2}$.
Although it is difficult to detect any well-defined 
trend of the correlation strength at $\Delta v \le 100$ km s$^{-1}$
in the two-point correlation functions,
the step optical depth correlation functions show 
a strong clustering at $v < 100$ km s$^{-1}$.
Instead of the individual pixels,
the two-point correlation function uses
the fitted line
lists from the Voigt profile fitting. The profile fitting
usually does not deblend saturated lines 
(higher $\tau$ and higher $N_\ion{H}{i}$) and is less
sensitive to fitting lines at 
small velocity separations.

The step optical depth correlation strength increases
as $z$ decreases, except
at $<\!z\!> \ =2.13$ at $\Delta v \le 50$ km s$^{-1}$.
The stronger correlation strength of the $<\!z\!> \ =2.13$
forest than the $<\!z\!> \ =1.98$ forest at 
$\Delta v \le 30$ km s$^{-1}$ 
is in part caused by the 5 strong lines 
clustered at 3960--4005 \AA\/.
Also the $<\!z\!> \ =2.13$ forest contains 
a void of $44\,h^{-1}$ Mpc.

The pixel-by-pixel correlation functions are
sensitive to the S/N and resolution of the data. However, the general
trend of increasing correlation strength with decreasing $z$
holds at the different S/N and resolution from the experiments
of degraded UVES spectra, although the degree of the correlation
strengths gets weaker as the S/N and resolution decrease.
With higher $\tau_\mathrm{min}$ values,
the correlation strengths get stronger,
but keep the same $z$-dependent trend. These 
results are in good agreement with
previous findings from the two-point velocity
correlation function: stronger forest lines are
more strongly correlated and the correlation strengths
increase with decreasing $z$ over a given column density range
(Cristiani et al. \cite{cri97}; Kim et al. \cite{kim97}).

\section{Discussion}

\subsection{The line number density 
per unit redshift of the Ly$\alpha$ forest}

For a given column density range, combined
HST and ground-based observations have provided 
evidence for
a change in the line number density evolution at 
$z \sim 1.7$: a rapid evolution at $z > 1.7$ and
a slow evolution at $z < 1.7$ (Impey et al. \cite{imp96};
Riediger et al. \cite{rie98};
Weymann et al. \cite{wey98}; Dav\'e et al. \cite{dav99}). 
These observations have led to a speculation of
two distinct populations in the Ly$\alpha$ forest: a rapidly
evolving population which dominates at higher $z$ and
a slowly evolving population which dominates at lower $z$.

Our results at $N_\ion{H}{i} = 10^{13.64-16}
\mathrm{cm}^{-2}$ suggest that the transition from
the stronger evolution to the weaker evolution in
$dn/dz$ occurs at $z \sim 1.2$ (\object{UM18}
fits in the picture and is not an outlier), rather than
at $z \sim 1.7$
as suggested by previous observations and numerical simulations. 
To be conservative, our
results show that $dn/dz$ 
at $N_\ion{H}{i} = 10^{13.64-16} 
\mathrm{cm}^{-2}$ continues to decrease at
a similar rate from $z = 4 \rightarrow 1.5$, with 
a suggestion of slowing down in the evolution towards lower $z$.

The physics of the Ly$\alpha$ forest at $z > 2$ is
determined mainly by the Hubble expansion and the ionizing background,
$J_\nu$ (or the \ion{H}{i} photoionization rate, 
$\Gamma_\ion{H}{i}$). 
If the forest is ``fixed" in comoving coordinates
for $q_0 = 0.5$ and $\Lambda = 0$, 
the observed number density of the Ly$\alpha$ forest is
proportional to $[(1+z)^{5} \, \Gamma_\ion{H}{i}^{-1}(z)]^{\beta-1} 
\, (1+z)^{-3/2}$,
where $\beta$ is from $f(N_\ion{H}{i}) \propto N_\ion{H}{i}^{-\beta}$. 
This implies that $dn/dz$ for a given column density
threshold decreases as $z$ decreases 
(Miralda-Escud\'e et al. \cite{mir96}; Dav\'e et al. \cite{dav99}). 
If we assume $\beta \sim 1.5$,
$dn/dz \propto (1+z)\,\Gamma_\ion{H}{i}^{-0.5}(z)$.
For a constant $\Gamma_\ion{H}{i}$, this is much lower than
the observed index of $dn/dz$, $2.19 \pm 0.27$,
suggesting structure evolution and/or  
$\Gamma_\ion{H}{i}$ evolution in $dn/dz$.

Recent numerical simulations 
suggest that a decrease of $\Gamma_\ion{H}{i}$ at $z < 2$
plays a more important role to change the slope in 
$dn/dz$ at $z \sim 1.7$ due to the decreasing 
QSO luminosity function at
$z < 2 $ (Dav\'e et al. \cite{dav99};
Riediger et al. \cite{rie98}; Theuns et al. \cite{the98};
Zhang et al. \cite{zha98}).
The discrepancy between our observations and simulations 
could be due to limited box sizes at $z < 2$ in 
most simulations (losing large-scale power),
to numerical resolutions (underestimating $\tau_\ion{H}{i}$
or the number of lines at lower $z$) or to incorrect 
$\Gamma_\ion{H}{i}$. 
If we take the results from most simulations
that $\Gamma_\ion{H}{i}$ is the main drive of the 
slope change in $dn/dz$, this discrepancy could simply indicate that
$\Gamma^{-1}_\ion{H}{i}$ in most simulations,
i.e. the QSO-dominated Haardt-Madau $\Gamma^{-1}_\ion{H}{i}$
(Haardt \& Madau \cite{haa96}), 
is underestimated at $z < 2$ and 
that $\Gamma^{-1}_\ion{H}{i}$ at $z < 2$
changes more slowly than a QSO-dominated 
$\Gamma^{-1}_\ion{H}{i}$, i.e.
there is a non-negligible contribution from galaxies at $z < 2$.

\subsection{The $z$-evolution of $b_{c}(N_\ion{H}{i})$}

Assuming a truncated Gaussian $b$ distribution with a lower 
$N_\ion{H}{i}$-independent
$b_{c}$, Kim et al. (\cite{kim97}) concluded that $b_{c}$
over $N_\ion{H}{i} = 10^{12.8-16} \ \mathrm{cm}^{-2}$
increases as $z$ decreases: 15 km s$^{-1}$ at $z \sim 3.7$, 17 km s$^{-1}$
at $z \sim 3.3$, 20 km s$^{-1}$ at $z \sim 2.9$, and 22--24 km s$^{-1}$
at $z \sim 2.3$. This result has been explained by an additional heating 
due to the on-going
\ion{He}{ii} reionization, although the high $b_{c}$ value at $z \sim 2.3$
does not agree with any theoretical explanations (Kim et al. \cite{kim97}; 
Haehnelt \& Steinmetz \cite{hae98};
Theuns et al. \cite{the00b}; Schaye et al. \cite{sch00}).

Subsequent studies on
the $z$-evolution of $b_{c}$ have led to contradictory results.
While $b_{c}$ is clearly dependent
on $N_\ion{H}{i}$ (Kirkman \& Tytler \cite{kir97}; Zhang et al. \cite{zha97}),
$b_{c} (N_\ion{H}{i})$ and the mean $b$ value at $<\!z\!> \ = 2.7$
(Kirkman \& Tytler \cite{kir97}) and at $<\!z\!> \ = 1.7$ 
(Savaglio et al. \cite{sav99})
does not show any noticeable difference compared with at $z > 3$.
Combining the observations with the numerical simulations,
Schaye et al. (\cite{sch00}) found
that $b_{c}$ at the fixed overdensity $\delta=0$,
$b_{\delta=0}$, increases from $z \sim 4.5$ ($b_{\delta=0}
\sim 14.5$ km s$^{-1}$) to $z \sim 3$ ($b_{\delta=0} \sim 
19.5 $ km s$^{-1}$)
due to \ion{He}{ii} reionization at $z \sim 3$
and then decreases from $z \sim 3$ ($b_{\delta=0} \sim 19.5 $ km
s$^{-1}$) to $z \sim 1.8$ ($b_{\delta=0} \sim 14 $ km s$^{-1}$).
Ricotti et al. (\cite{ric00}) also found a similar increase in 
$b_{\delta=0}$ at $z \sim 3$, although their $b_{\delta=0}$
at $z > 2.8$ and at $z < 2.8$ can be considered to be constant 
at $b_{\delta=0} \sim 14.5$ km s$^{-1}$ and at 
$b_{\delta=0} \sim 21$ km s$^{-1}$, respectively.
On the other hand, adopting a slightly different approach for
identifying absorption lines instead of the Voigt profile fitting,
McDonald et al. (\cite{mc00}) found that there is
no $b_{c}$ evolution over $ 2.1 < z < 4.4$ at 
the slightly higher overdensity $\delta=0.4$.

\begin{figure}
\resizebox{\hsize}{!}{\includegraphics{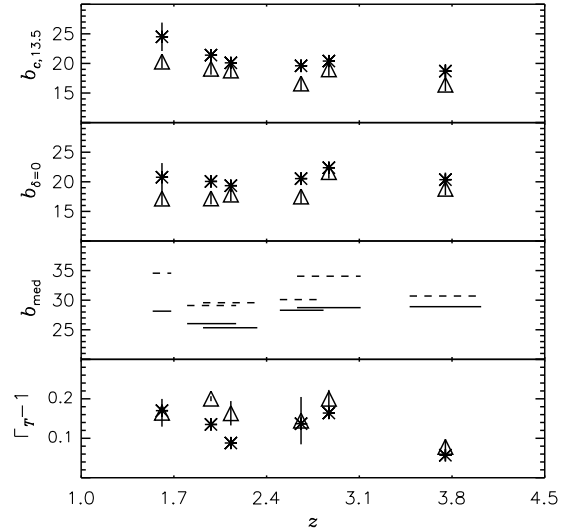}}
\caption{The $z$-evolution of $b_{c, \mathrm{13.5}}$,
$b_{\delta=0}$, median $b$ and $(\Gamma_{T}-1)$
from the two power law fits in Sect. 4.3.
Stars and triangles represent 
the parameters measured from the iterative power law fit 
and from the smoothed
$b$ distribution, respectively. Vertical bars
represent 1$\sigma$ errors. In the third panel,
error bars in the x-axis represent the $z$ ranges over which 
the median $b$ values were estimated. Solid 
lines represent the median
$b$ values over $N_\ion{H}{i} = 10^{13.1-14} \ \mathrm{cm}^{-2}$,
while dotted lines represent the median
$b$ values over $N_\ion{H}{i} = 10^{13.8-16} \ \mathrm{cm}^{-2}$.}
\label{fig_gamma1}
\end{figure}

Fig.~\ref{fig_gamma1} shows
the cutoff $b$ values at the fixed column density
$N_\ion{H}{i} = 10^{13.5} \ \mathrm{cm}^{-2}$, $b_{c, \mathrm{13.5}}$,
from the two
power law fits in Sect. 4.3 as a function of $z$.
Keep in mind that $b_{c} (N_\ion{H}{i})$
from the iterative power law fit is an upper limit, while
$b_{c} (N_\ion{H}{i})$ from the smoothed $b$ distribution
is a lower limit. 
The $b_{c, \mathrm{13.5}}$ value
shows a slight increase with decreasing $z$
from both power law fits, with a possible local $b_{c, \mathrm{13.5}}$
maximum at $z \sim 2.9$ (with the caveat that the line list of
\object{Q0302--003} is generated by a different fitting program
with respect to the other line lists. This could introduce an
artificial result at $z \sim 2.9$ as shown in Sect. 5.4, although
its redshift range suggests an influence of additional heating
if \ion{He}{ii} reionization does occur at $z \sim 3$).
When all the values from both
fits are averaged, 
$b_{c, \mathrm{13.5}} = 17.6 \pm 1.6$ at $z \sim 3.75$ is 
smaller than $b_{c, \mathrm{13.5}} = 19.9 \pm 1.2$
at $z \sim 2.1$, but the difference is significant only at the 
$1.44\sigma$ level.

In simulations, $b_{c}$ is usually measured at a fixed
overdensity $\delta$ rather than at a fixed column density. Translating
an overdensity into the corresponding column density is not
trivial and depends on many uncertain parameters, such as 
the ionizing background and the reionization history. 
If the simple law between
$\delta$ and $N_\ion{H}{i}$ by Dav\'e 
et al. (\cite{dav99})\footnote{It should be noted that Dav\'e
et al. (\cite{dav99}) assume the QSO-dominated Haardt-Madau UV
background without \ion{He}{ii} reionization. No other
scaling laws between $N_\ion{H}{i}$
and $\delta$ as a function of $z$ from simulations under
different UV backgrounds are found in the literature.}
is assumed,
then $\delta$ becomes:
\begin{equation}
N_\ion{H}{i} \sim \left[0.05\, (1+\delta) \, 
10^{0.4 \,z}\right]^{1.43} \times 10^{14} \ 
\mathrm{cm}^{-2}. 
\label{eq1}
\end{equation}
For $\delta=0$, the corresponding $N_\ion{H}{i}$ is
$1.92 \times 10^{14}$, $6.04 \times 10^{13}$, $4.56 \times 10^{13}$,
$2.28 \times 10^{13}$, $1.87 \times 10^{13}$ and
$1.15 \times 10^{13} \ \mathrm{cm}^{-2}$ at
$<\!z\!> \ =$ 3.75, 2.87, 2.66, 2.13, 1.98 and 1.61,
respectively. Although this conversion does not include
the effects of the \ion{He}{ii} reionization,
we assume that it is correct at least
on a relative scale at $z \sim 3.75$
and at $z \sim 2.1$, where the \ion{He}{ii} reionization 
would not affect the temperature of the IGM as strongly as 
at $z \sim 3$.\footnote{ 
Note that Ricotti et al. (\cite{ric00})
found that $\delta=0$ corresponds to $N_\ion{H}{i} = 5.62 \times
10^{12} \ \mathrm{cm}^{-2}$ at $z = 2.85$ with
\ion{He}{ii} reionization at $z \sim 3$, almost a factor of
10 smaller than the column density calculated by Eq.~\ref{eq1}.
This discrepancy indicates the importance of using the correct ionizing
background in simulations to constrain the temperature
of the IGM.}
Therefore, we only discuss the relative behavior of $b_{\delta=0}$
as a function of $z$, in particular,
our $b_{\delta=0}$ at $z \sim 2.1$
with other results mentioned above.

For the iterative power law fit (the power law fit to the
smoothed $b$ distribution), the $b$ value at $\delta=0$, $b_{\delta=0}$,
is 20.8 (17.2),
20.1 (17.2), 19.4 (17.8), 20.5 (17.5), 22.4 (21.6)
and 20.4 (18.8) km s$^{-1}$ at 
$<\!z\!> \ =$ 3.75, 2.87, 2.66, 2.13, 1.98 and 1.61, respectively.
In the second panel of Fig.~\ref{fig_gamma1},
$b_{\delta=0}$ is fairly
constant with $z$ as $b_{\delta=0} \sim$ 17--20 km s$^{-1}$,
with a 
possible local maximum $b_{\delta=0} \sim$ 22 km s$^{-1}$
at $z \sim 2.9$. 
The observed behavior of $b_{\delta=0}$ is qualitatively
in agreement with the results from McDonald et al. (\cite{mc00}).
While the observations agree with the fairly constant
$b_{\delta=0}$ at $z < 3$ derived by Ricotti et al. (\cite{ric00}),
they do not show the abrupt increase of $b_{\delta=0}$ 
across $z \sim 3$ as large as 
$\sim 7$ km s$^{-1}$ found by Ricotti et al.
The observations at $z \sim 3.75$ and at $z \sim 2.1$
agree with the results by Schaye et al. (\cite{sch00})
which show similar $b_{\delta=0}$ at $z \sim 3.75$
and at $z \sim 2.1$. In addition,
the observations do not show
a strong decrease of $b_{\delta=0}$ from $z \sim 3$ to $z \sim 2$
as large as $\sim 4$ km s$^{-1}$ as found by Schaye et al. (\cite{sch00}).
However, note that, considering the large error bars of Schaye et al.
(\cite{sch00}), the significance of the decrease of $b_{\delta=0}$
from $z \sim 3$ to $z \sim 2$ is not very strong and their
result is not in disagreement with ours.
It should also be recalled that we are
using the scaling law between $\delta$ and $N_\ion{H}{i}$
estimated from the QSO-dominated 
Haardt-Madau $\Gamma_\ion{H}{i}$. If this
QSO-dominated UV background is underestimated at $z < 2$
as suggested by the evolution of the absorption line number density,
the {\it actual} $b_{\delta=0}$ at $z < 2.4$ can be higher than
$b_{\delta=0}$ in Fig.~\ref{fig_gamma1}.

The third panel of Fig.~\ref{fig_gamma1} shows the median
$b$ values as a function of $z$ measured for two column density ranges:
$N_\ion{H}{i} = 10^{13.1-14} \mathrm{cm}^{-2}$ and
$N_\ion{H}{i} = 10^{13.8-16} \mathrm{cm}^{-2}$.
It is rather difficult to interpret 
the $z$-dependence of the median $b$
values. It could be constant at $1.5 < z < 4$ with a small cosmic
variance at $z \sim 2$. On the other hand, it could be decreasing
with $z$ at $z < 3.1$, if we discard the median $b$ values at
$<\!z\!> \ = 1.61$, based on a small number of absorption lines.
Although simulations correctly predict the shape
of the observed $b$ distributions,
the predicted median $b$ values are typically 5--10 km s$^{-1}$
smaller than the observed ones at all $z$
(Bryan \& Machacek \cite{bry00}; Machacek et al. \cite{mac00};
Theuns et al. \cite{the00b}). 

The bottom panel of Fig.~\ref{fig_gamma1} shows the power law slope
of the $N_\ion{H}{i}$--$b_{c}$ distribution,
$(\Gamma_{T}-1)$, as a function of $z$ (see Eq.~\ref{eq0}).
No particular trend is apparent and $(\Gamma_{T}-1)$ 
shows very little evolution at $z < 3.1$. Note that the
lower $(\Gamma_{T}-1)$ at $<\!z\!> \ = 3.75$ is in part due to
the lack of lines with $b < 15$ km s$^{-1}$ and $N_\ion{H}{i}
\le 10^{13.4} \mathrm{cm}^{-2}$ (Fig.~\ref{fig_test1}). 
When $(\Gamma_{T}-1)$ is averaged over all the measured values,
$(\Gamma_{T}-1) = 0.16 \pm 0.03$ at $z < 3.1$ is larger than
$(\Gamma_{T}-1) = 0.07 \pm 0.02$ at $<\!z\!> \ = 3.75$.
Due to the lower $(\Gamma_{T}-1)$ at $<\!z\!> \ = 3.75$, it
also seems clear from Fig.~\ref{fig_test1} that
gas at lower overdensities is
cooler at $z < 3.1$ than at higher $z$
(keep in mind that a fixed column density corresponds
to a larger gas overdensity as $z$ decreases due to the Hubble
expansion).

If we assume Eq.~\ref{eq1} again and $T\sim 59.2 \, b^{2}$ for
thermally broadened lines, $(\gamma_{T}-1) \sim
2.857 \, (\Gamma_{T}-1)$.
When this simple conversion law is assumed,
$(\gamma_{T}-1) = 0.46 \pm 0.10$ at $z < 3.1$ and
$(\gamma_{T}-1) = 0.18 \pm 0.05$ at $z \sim 3.75$. 
These $(\gamma_{T}-1)$ could be considered to be
consistent with the results by McDonald et al.
(\cite{mc00}) within their error bars, although their
error bars at $z =3.9$ ($(\gamma_{T}-1) = 0.42 \pm 0.45$) 
and at $z=3$ ($(\gamma_{T}-1) = 0.30 \pm 0.30$) are rather large.
Their $(\gamma_{T}-1)=0.5 \pm 0.15$ at $z \sim 2.4$ 
agrees with our $(\gamma_{T}-1) = 0.46$.
Our $(\gamma_{T}-1)$ is marginally in agreement 
with the results from Ricotti et al. (\cite{ric00})
and from Schaye et al. (\cite{sch00}). 
Their $(\gamma_{T}-1)$ values are lower
at $z \sim 2$ and higher at $z \sim 3.7$, but within the error
bars. 

\subsection{The optical depth distribution function}

\begin{figure}
\resizebox{\hsize}{!}{\includegraphics{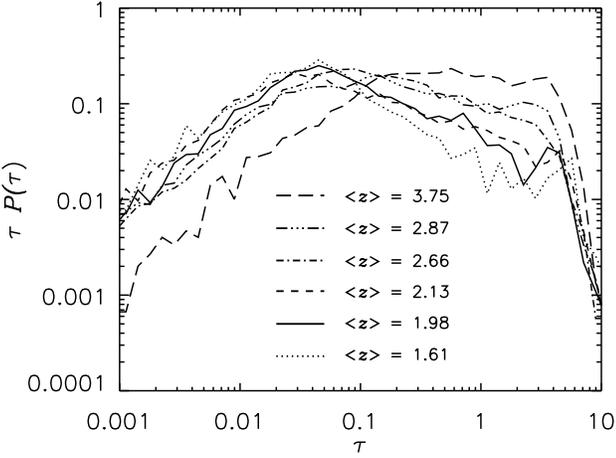}}
\caption{The $z$-evolution of $\tau P(\tau)$ as a function
of $\tau$. The $\tau P(\tau)$ values are
calculated from the observed spectra at $z < 2.4$ and
from the spectra generated with noise at $z > 2.4$.}
\label{fig_zhang}
\end{figure}

While the one-point function of the flux is more closely related
to observations, the one-point function of the optical depth
is usually calculated in simulations (Zhang et al. \cite{zha98};
Machacek et al. \cite{mac00}). In the simulation by
Machacek et al. (\cite{mac00}), the \ion{H}{i} optical depth $\tau$ at which 
the maximum $\tau P(\tau)$ occurs, $\tau_{\mathrm{max}}$, is 
$\tau \sim 0.8$ at $z=4$, $\tau \sim 0.09$ at $z=3$,
and $\tau \sim 0.013$ at $z=2$.
Fig.~\ref{fig_zhang} shows $\tau P(\tau)$ as a function of $\tau$,
which is calculated from the observed spectra at $z < 2.4$
and the spectra generated
with noise at $z > 2.4$. The $\tau_{\mathrm{max}}$ value is 
$\tau \sim 0.6$ 
at $<\!z\!> \ = 3.75$, 
$\tau \sim 0.13$ at $<\!z\!> \ = 2.87$, and
$\tau \sim 0.08$ at $<\!z\!> \ = 2.66$, respectively.
The observed $\tau_{\mathrm{max}}$ shows a behavior similar
to the simulated results by Machacek et al. (\cite{mac00}).
However, $\tau_{\mathrm{max}}$ converges to $\tau \sim 0.04$ 
at $z < 2.4$, not showing any $z$-dependence. Also there is
no $z$-dependence of $\tau P(\tau)$ at $\tau \le 0.04$ and at
$z < 3.1$.

The optical depth $\tau \sim 0.04$ corresponds to $F \sim 0.96$, while 
$\tau \sim 0.013$ corresponds to
$F \sim 0.99$, almost to the continuum level. 
As $z$ decreases, the number of pixels
with $F =$ 0.96--0.99 increases. These pixels are
noise-dominated by the limited S/N and the continuum
fitting uncertainty.
Therefore, instead of showing
the expected $z$-dependence of $\tau_{\mathrm{max}}$ and
$\tau P(\tau)$, the observed
$\tau_{\mathrm{max}}$ and $\tau P(\tau)$ approach to
an asymptotic $\tau_{\mathrm{max}} \sim 0.04$ value at $z < 2.4$
and an asymptotic $\tau P(\tau)$ at
$\tau \le 0.04$ and at $z < 3.1$, respectively.

At $0.1 < \tau < 3$ (or $0.05 < F < 0.9$), 
the observed $\tau P(\tau)$ is simply
a different way of viewing the one-point function of the flux.
As $z$ decreases, $\tau P(\tau)$ decreases due to the expansion
of the universe. At $\tau > 4$ (or $F < 0.05$), $\tau P(\tau)$
starts to converge again since it typically samples saturated regions, again
dominated by noise.

\subsection{The baryon density $\Omega_{b}$}

We derived the baryon density, $\Omega_{b}$, from two
properties of the Ly$\alpha$ forest, $\overline{\tau}_\ion{H}{i}$
(including all the Ly$\alpha$ forest regardless of $N_\ion{H}{i}$)
and $P(F)$ (only for 
the forest with $N_\ion{H}{i} \le 10^{16} \mathrm{cm}^{-2}$). 

If $\overline{\tau}_\ion{H}{i} \ \le 1.24$, the lower limits on the
baryon density become
\begin{eqnarray*}
\Omega_{b}^{\overline{\tau}_\ion{H}{i}} \ge 5.1 \times 10^{4} \,
(1-\exp(- \overline{\tau}_\ion{H}{i}))\,
(1+z)^{-2.5} \,
(1 + \Omega_\mathrm{0}z)^{0.25}
\end{eqnarray*}
\begin{equation}
\ \ \ \ \ \ \ \ \ \ \times h^{-1.75}
T_{0}^{0.35} \,
\Gamma_\ion{H}{i}^{0.5},
\label{eq2}
\end{equation}
where all the symbols have their usual meanings in this study
(Weinberg et al. \cite{wei97}).

Table~\ref{Tab9} lists the lower limits on $\Omega_{b}$
at different $z$, together with the parameter values
used for calculating
$\Omega_{b}^{\overline{\tau}_\ion{H}{i}}$.
It should be noted that
the temperature used to 
calculate Eq.~\ref{eq2} is smaller than the 
values derived in Sect. 6.2, but in the present discussion 
we are only interested in the lower
bounds on $\Omega_{b}$. These lower limits on
$\Omega_{b}^{\overline{\tau}_\ion{H}{i}}$ are consistent with
$\Omega_{b} =0.0125 \, h^{-2}$ from the Big Bang
nucleosynthesis analysis (Copi et al. \cite{cop95}). These
values also indicate that about 90\% of all baryons reside
in the Ly$\alpha$ forest at $1.5 < z < 4$.

The lower bounds on the density parameter from the one-point
function, $\Omega_{b}^{\mathrm P(F)}$,
are given by Weinberg et al. (\cite{wei97}) as 
\begin{eqnarray*}
\Omega_{b}^{\mathrm P(F)} & \ge & 0.021 h^{-3/2} \left(
\frac{[\int^{1}_{0} (-\ln F)^{1/\beta_{\gamma}} P(F)\,
dF/F]^{\beta_{\gamma}/2}}{0.70}
\right)
\end{eqnarray*}
\begin{eqnarray*}
\ \ \ \ \ \ \ \ \ \ \times \left(\frac{4}{1+z}\right)^{3}
\left(\frac{H(z)/H_\mathrm{0}}{5.51}\right)^{1/2}
\left( \frac{T_\mathrm{0}}{10^{4}
{\mathrm K}}
\right )^{0.35}
\end{eqnarray*}
\begin{equation}
\ \ \ \ \ \ \ \ \ \ \times \left( \frac{\Gamma}{10^{-12}
\ \mathrm{sec}^{-1}} \right)^{1/2},
\label{eq3}
\end{equation}
where $\beta_{\gamma} \equiv 
(2-0.7\gamma_{T})^{-1} =1.6-1.8$,
where $\gamma_{T}$ is the power law index of the equation of state
(Weinberg et al. \cite{wei97}).

Table~\ref{Tab9} lists the lower bounds on 
$\Omega_{b}^{\mathrm{P(F)}}$ along with the
parameter values used to calculate $\Omega_{b}^{\mathrm{P(F)}}$.
The $\Omega_{b}^{\mathrm{P(F)}}$ values are larger than 
$\Omega_{b}^{\overline{\tau}_\ion{H}{i}}$ since
$\overline{\tau}_\ion{H}{i}$ is not the true mean \ion{H}{i}
opacity, but the effective opacity which underestimates
the true opacity when absorption lines become saturated.
The lower $\Omega_{b}$ limits from
$P(F)$ are about a factor
of $1.6 \, h^{0.5}$ larger than 
the Big Bang nucleosynthesis analysis,
$\Omega_{b} =0.0125 \, h^{-2}$. 

Our new lower bounds on $\Omega_{b}$ are a factor of
1.5 smaller than some of the previous results,
$\Omega_\mathrm{b} = 0.017$--$0.03 \, 
h^{-2}$ (Rauch et al. \cite{rau97};
Zhang et al. \cite{zha98}; Burles et al. \cite{bur99}; 
Kirkman et al. \cite{kir00};
McDonald et al. \cite{mc00}),
but still consistent with them within the error bars.
However, our lower $\Omega_\mathrm{b}$ bounds are
not consistent with the derived $\Omega_{b} =
0.005 \, h^{-2}$--$0.01 \, h^{-2}$ from the high D/H measurements
(Songaila et al. \cite{son94}; Rugers \& Hogan \cite{rug96}).

\begin{table}
\caption[]{The lower bounds on $\Omega_{b}$}
\label{Tab9}
\begin{tabular}{cccc}
\hline
\noalign{\smallskip}
$z$ & $\overline{\tau}_\ion{H}{i}$ & $\Omega_{b}^{\mathrm{a}}$ &
$\Omega_{b}^{\mathrm{P(F),b}}$ \\
& & ($\times h^{-1.75}$) & ($\times h^{-1.5}$) \\
\noalign{\smallskip}
\hline
\noalign{\smallskip}
1.61 & 0.086 & $\ge 0.010$ & $\ge 0.020$ \\
1.98 & 0.161 & $\ge 0.016$ & $\ge 0.021$ \\
2.13 & 0.131 & $\ge 0.012$ & $\ge 0.021$ \\
2.66 & 0.234 & $\ge 0.014$ & $\ge 0.020$ \\
2.87 & 0.275 & $\ge 0.015$ & $\ge 0.020$ \\
3.75 & 0.733 & $\ge 0.013$ & $\ge 0.013$ \\
\noalign{\smallskip}
\hline
\end{tabular}
\begin{list}{}{}
\item[$^{\mathrm{a}}$] For $T = 6\,000$ K, $\Gamma =
0.9 \times 10^{-12} \ \mathrm{sec}^{-1}$ (at $z \sim 1.6$), 
$1.4 \times 10^{-12} \ \mathrm{sec}^{-1}$ (at $z \sim $2--3),
$0.64 \times 10^{-12} \ \mathrm{sec}^{-1}$ (at $z \sim 3.75$),
and $\Omega_{0}=1$
(Haardt \& Madau \cite{haa96};
Hui \& Gnedin \cite{hui97a}; Weinberg et al. \cite{wei97}).
\item[$^{\mathrm{b}}$] For $T = 6\,000$ K, $\Gamma =
0.9 \times 10^{-12} \ \mathrm{sec}^{-1}$ (at $z \sim 1.6$), 
$1.4 \times 10^{-12} \ \mathrm{sec}^{-1}$ (at $z \sim $2--3),
$0.64 \times 10^{-12} \ \mathrm{sec}^{-1}$ (at $z \sim 3.75$),
$\Omega_{0}=1$,
$\beta=1.58$, and
$H(z)=H_{\rm 0} \Omega_{\rm 0}^{\frac{1}{2}} (1+z)^{\frac{3}{2}}$
(Haardt \& Madau \cite{haa96}; 
Hui \& Gnedin \cite{hui97a}; Weinberg et al. \cite{wei97}).
\end{list}
\end{table}

\section{Conclusions}
We have analyzed 
the properties of low column density Ly$\alpha$ forest
clouds ($N_\ion{H}{i} = 10^{12.5 -16} \mathrm{cm}^{-2}$) toward 3 QSOs at
$1.5 < z < 2.4$, using high resolution ($R \sim 45\,000$),
high S/N ($\sim$ 25--40) VLT/UVES data.
Combined with other high-resolution observations from the literature
at $z > 2.4$, 
we have studied the evolution of the Ly$\alpha$ forest 
at $1.5 < z < 4$.
Two parallel analyses have been applied to the datasets:
the traditional Voigt profile fitting analysis
and a statistical measure of the transmitted flux.
We find that the general conclusions from both analyses are in
good agreement. Although the results are limited by the relatively
small number of lines of sight,
we find the following 
properties and trends in the $z$-evolution of the Ly$\alpha$ forest:

1) The differential density distribution function of the lower column
density forest ($N_\ion{H}{i} = 10^{12.5-14} \ \mathrm{cm}^{-2}$)
does not evolve very strongly, $f(N_\ion{H}{i}) \propto
N_\ion{H}{i}^{-\beta}$, at $1.5 < z < 4$, with 
$\beta \sim 1.4$--1.5 and with an
indication of an increasing $\beta$ to $\beta \sim 1.7$ 
at $z < 1.8$. 
The higher column density forest ($N_\ion{H}{i} > 10^{14} \
\mathrm{cm}^{-2}$)
disappears rapidly with decreasing $z$.
The observed slopes of $f(N_\ion{H}{i})$ for various
column density ranges 
are considerably flatter than numerical predictions
for the same column density thresholds.
The same conclusions are drawn from the 
one-point function and two-point function of the flux.

2) The line number density of the Ly$\alpha$ forest with
$N_\ion{H}{i} = 10^{13.64-16} \ \mathrm{cm}^{-2}$ 
decreases continuously from
$z \sim 4$ to $z \sim 1.5$, 
$dn/dz \propto (1+z)^{2.19 \pm 0.27}$, without
showing any flattening in $dn/dz$ at $1.5 < z < 4$. 
For the lower column density range at
$N_\ion{H}{i} = 10^{13.1-14} \ \mathrm{cm}^{-2}$, the 
number density evolution becomes weaker:
$dn/dz \propto (1+z)^{1.10 \pm 0.21}$.
The line counts as a function of the filling factor
show a negligible $z$-dependence.
These results strongly suggest that the main
drive in the evolution of the forest at $z > 1.5$ is the Hubble
expansion and that $\Gamma_\ion{H}{i}$ changes more slowly than
a QSO-dominated background at $z < 2$, 
suggesting a contribution from galaxies
to the UV background at $z < 2$. 
When combined with the results from the HST QSO absorption
line key project at $0 < z < 1.5$ with the
same column density threshold, there
is evidence for a slope change in $dN/dz$ at $z \sim 1.2$.

3) Deriving the cutoff $b_c (N_\ion{H}{i})$ 
as a function of $z$ depends strongly on
the methods used and the number of available lines. 
However,
the cutoff $b$ parameter at the fixed column density
$N_\ion{H}{i} = 10^{13.5} \ \mathrm{cm}^{-2}$,
$b_{c, \mathrm{13.5}}$, shows a weak increase with
decreasing $z$, with a possible local $b_{c, \mathrm{13.5}}$
maximum at $z \sim 2.9$. Despite being substantially uncertain
due to the uncertain conversion from the observable
parameters $b$ and $N_\ion{H}{i}$ to the theoretical
parameters $T$ and $\delta$, 
the cutoff $b$ value at the mean gas density, 
$b_{\delta=0}$, is
fairly constant with $z$ as $b_{\delta=0} \sim 18$ km s$^{-1}$,
with a possible local $b_{c}$ maximum at $z \sim 2.9$.
The observed slopes of $b_{c} (N_\ion{H}{i})$ 
do not show any well-defined $z$-dependence except for a
flatter slope at $<\!z\!> \ = 3.75$ than at $z < 3.1$, 
possibly due
to the severe blending of 
lines with low-$b$ and low-$N_\ion{H}{i}$ at higher
$z$.

4) The velocity correlation function and the step optical-depth
correlation function confirm that stronger
lines are more clustered than weaker lines and that the
correlation strength increases as $z$ decreases.

5) The mean \ion{H}{i} opacity is well approximated by a single
power law at $1.5 < z < 4$, $\overline{\tau}_\ion{H}{i} 
\propto (1+z)^{3.43 \pm 0.17}$, 
without showing any flattening towards higher $z$. The
significant scatter in the mean \ion{H}{i} opacity is likely to be
caused by continuum fitting uncertainty and by an 
inclusion/exclusion
of high-column-density \ion{H}{i} systems 
($N_\ion{H}{i} \ge 10^{17} \ \mathrm{cm}^{-2}$)
towards different lines of sight.

6) The lower limit on the baryon density 
derived both 
from the \ion{H}{i} opacity and the one-point function
of the flux,
$\Omega_\mathrm{b}^{\mathrm{Ly\alpha}} \, 
\sim 0.013 \, h^{-1.75}$, suggests that most
baryons (over 90\%) reside in the forest at $1.5 < z < 4$.
The contribution to $\Omega_\mathrm{b}$ from the Ly$\alpha$ forest
does not change
much with $z$ at $1.5 < z < 4$.

\begin{acknowledgements}
We are indebted to all the people involved in the conception, construction and
commissioning of UVES and UT2 for the quality of the data used in this paper,
obtained in the first weeks of operation of the instrument.
TSK would like to express her gratitude to 
Vanessa Hill and Sebastian Wolf for their generous help
on using the MIDAS data reduction software, to Michael Rauch
and Bob Carswell on
using VPFIT and to Aaron Evans, Pamela Bristow and our editor
Jet Katgert on their
careful reading of the manuscript. We are also in debt to
Jacqueline Bergeron and Martin Haehnelt 
for insightful discussions.
We are also deeply grateful to the
referee, Alain Smette, for his
careful reading of the manuscript and for helpful discussions to improve
our study. 
This work has been conducted with partial support by the Research
Training  Network "The Physics of the Intergalactic
Medium" set up by the European
Community under the contract HPRN-CT2000-00126 RG29185 and by
ASI through contract ARS-98-226.

\end{acknowledgements}

\end{document}